\documentclass[fleqn,usenatbib]{mnras}

\usepackage{newtxtext,newtxmath}

\usepackage[T1]{fontenc}

\DeclareRobustCommand{\VAN}[3]{#2}
\let\VANthebibliography\thebibliography
\def\thebibliography{\DeclareRobustCommand{\VAN}[3]{##3}\VANthebibliography}

\usepackage{graphicx}	
\usepackage{amsmath}	
\usepackage{amssymb}	
\usepackage{subcaption}
\usepackage{xcolor}
\definecolor{ao}{rgb}{0.0, 0.5, 0.0}
\definecolor{bv}{rgb}{0.54, 0.17, 0.89}
\definecolor{r}{rgb}{0.8, 0.0, 0.0}
\definecolor{notegreen}{rgb}{0.235,0.651,0.282}

\newcommand{\oiii}{[\mbox{O\,\sc{iii}}]}
\newcommand{\oiiihb}{[\mbox{O\,\sc{iii}}]+$\mathrm{H\,\beta}$}
\newcommand{\EWoiiihb}{$W_{\lambda}$([\mbox{O\,\sc{iii}}]+$\mathrm{H\,\beta}$)}
\newcommand{\Lya}{$\mathrm{Ly\alpha}$}
\newcommand{\xiion}{$\xi_\mathrm{ion}$}
\newcommand{\fesclyc}{$f_\mathrm{esc}^\mathrm{LyC}$}

\title[\mbox{The redshift evolution of the $\mathrm{[O\,}$\sc{iii}]}$+\mathrm{H}\beta$ equivalent width distribution]{The evolution of \mbox{[O\,\sc{iii}]}$+\mathrm{H}\,\beta$ equivalent width from $\mathbf{z\simeq3-8}$: implications for the production and escape of ionizing photons during reionization}
\author[R. Begley et al.]{R. Begley$^{1}$\thanks{E-mail:rbeg@roe.ac.uk}, R. J. McLure$^{1}$, F. Cullen$^{1}$, D. J. McLeod$^{1}$, J. S. Dunlop$^{1}$, A. C. Carnall$^{1}$, T. M. Stanton$^{1}$, \and A. E. Shapley$^{2}$, R. Cochrane$^{1}$, C. T. Donnan$^{1}$, R.S. Ellis$^{3}$, A. Fontana$^{4}$, N. A. Grogin$^{5}$, A. M. Koekemoer$^{5}$\\\\
$^{1}$Institute for Astronomy, University of Edinburgh, Royal Observatory, Edinburgh EH9 3HJ, UK\\
$^{2}$Department of Physics \& Astronomy, University of California, Los Angeles, 430 Portola Plaza, Los Angeles, CA 90095, USA\\
$^{3}$Department of Physics \& Astronomy, University College London, Gower St., London WC1E 6BT, UK\\
$^{4}$INAF - Osservatorio Astronomico di Roma, via di Frascati 33, 00078 Monte Porzio Catone, Italy\\
$^{5}$Space Telescope Science Institute, 3700 San Martin Drive, Baltimore, MD 21218, USA}
\date{Accepted 2025 February 03. Received 2025 January 31; in original form 2024 October 14}

\pubyear{2025}

\begin{document}
\label{firstpage}
\pagerange{\pageref{firstpage}--\pageref{lastpage}}
\maketitle

\begin{abstract}
    Accurately quantifying the ionizing photon production efficiency ({\xiion}) of $z\gtrsim6$ star-forming galaxies (SFGs) is
necessary to fully understand their contribution to reionization. In this study, we investigate the ionizing properties of
$N=279$ SFGs selected at $z\simeq6.9-7.6$ from two of the largest JWST Cycle-1 imaging programmes; PRIMER and JADES.
We use \textsc{bagpipes} to consistently infer the equivalent widths ($W_{\lambda}$) of their {\oiiihb} emission lines and their physical properties. To supplement this sample, we measure {\EWoiiihb} photometrically for $N=253$
$z_{\mathrm{spec}}=3.2-3.6$ SFGs selected from the VANDELS
spectroscopic survey. Comparing these samples, we find a strong apparent redshift evolution in
their median {\EWoiiihb}, increasing
from {\EWoiiihb $=380\pm30$\AA} in VANDELS to {\EWoiiihb$=540\pm25$\AA} in PRIMER$+$JADES.
Concentrating on the JWST sample ($z\gtrsim7$), we find that {\EWoiiihb} correlates with stellar mass and UV luminosity, with high-mass, $M_{\mathrm{UV}}$-faint galaxies producing systematically weaker emission lines.
Moreover, we discover a departure from the standard log-normal shape of the {\EWoiiihb} distribution, with a more pronounced tail towards lower {\EWoiiihb}, consistent with increasingly bursty star formation. Using {\EWoiiihb} as a proxy for {\xiion}, and UV spectral slope as a proxy for Lyman-continuum escape ({\fesclyc}), we uncover a minority of galaxies with high {\xiion} {\it and} {\fesclyc} (e.g., $\mathrm{log}(\xi_\mathrm{ion}/\mathrm{erg^{-1}Hz})\simeq25.6$ and {\fesclyc$\simeq 0.15$}). However, we find the ionizing photon budget at $z\gtrsim7$ is dominated
by galaxies with more moderate output, close to the median values of {$\mathrm{log}(\xi_\mathrm{ion}/\mathrm{erg^{-1}Hz})\simeq25.3$} and {\fesclyc}$\simeq 0.05$. Our results are consistent with estimates for the number of ionizing photons required to power reionization at $z\gtrsim7$, with no evidence for over or under-production.

\end{abstract}

\begin{keywords}
galaxies: high-redshift - galaxies: evolution - cosmology: dark ages, reionization, first stars
\end{keywords}



\section{Introduction}\label{sec:intro}
During the Epoch of Reionization (EOR), Hydrogen ionizing (i.e., Lyman continuum; LyC) photons permeated through the intergalactic medium (IGM), driving its transition from entirely neutral to fully ionized \citep{robertson+15,robertson+23}. Evidence from the {\Lya} forest of distant quasars point to this epoch ending at $z\simeq5.5$ \citep{fan+06,goto+21,bosman+21}, while \citet{planck20} measurements of the electron scattering optical depth ($\tau$) suggest an `instantaneous’ reionization midpoint of $z_\mathrm{re}=7.68\pm0.79$. Nonetheless, the exact timeline and topology of the reionization process remains uncertain \citep{becker+15,mason+18,garaldi+22}.
    
The demographics of the sources of ionizing photons plays a key role in dictating the overall progression of reionization \citep[e.g.,][]{robertson+15,mason+19,dawoodbhoy+23}. Early quasars likely only played a minor role in sustaining the LyC photon budget required to drive reionization due to their relative scarcity at high redshift \citep{aird+15,kulkarni+19,maiolino+23,matsuoka+23,trebitsch+23}. On the other hand, measurements of the UV luminosity function at $z>5$ \citep{bouwens+15,bowler+20,harikane+21,donnan+22,mcleod+24} indicate the presence of a large population of star-forming galaxies (SFGs) during the EOR, particularly at faint luminosities ($M_\mathrm{UV}\gtrsim-18$), as a result of relatively steep faint-end slopes \citep[e.g., $\alpha\lesssim-2$][]{finkelstein+15}.
Although a consensus has been reached regarding the dominant role of early SFGs in producing the bulk of the ionizing photon budget \citep{chary+16,robertson+23}, questions about the properties of these galaxies remain.\\

Two key galaxy properties interlinked with the EOR timeline and topology are 1.) their ionizing photon production rates, often quantified as the ionizing photon production efficiency {\xiion} ($\equiv N(\mathrm{H}^0)/L_\mathrm{UV}$), which is the number of LyC photons produced per unit UV luminosity, and 2.) {\fesclyc}, the fraction of photons produced that then escape into the surrounding IGM \citep[e.g.,][]{robertson+15,finkelstein+19,mason+19}. A number of state-of-the-art radiation-hydrodynamic simulations provide supporting evidence for a `democratic' reionization process, whereby the faint but numerous population of $M_\mathrm{UV}\gtrsim-18$ galaxies dominates the overall LyC photon budget \citep{lewis+22,rosdahl+22}.
This scenario is in contrast to models suggesting reionization is driven by the rarer, UV-luminous `oligarchs' \citep[e.g.,][]{naidu+20} or, alternatively, by a small subset of the brightest {\Lya} emitters with high {\xiion} and {\fesclyc} \citep[e.g., see;][]{matthee+22,naidu+22a}.\\

However, each of these models can be made consistent with the reionization history inferred from constraints on the evolution in the global neutral Hydrogen fraction \citep[e.g.,][]{mcgreer+15,hoag+19,mason+19}. It is therefore clear that deciphering the relative contributions of different galaxy sub-populations remains an open debate, and that a better understanding of {\xiion} and {\fesclyc} across these populations is required.\\

Direct measurements of {\fesclyc} are restricted to $z\lesssim4$ on account of the increasing opacity of the IGM to LyC photons at higher redshifts \citep{madau+95,inoue+14}.  Studies based on deep \textit{U-}band imaging and spectroscopy have shown SFGs at $z=3-4$ have $f_{\mathrm{esc}}^{\mathrm{LyC}}\simeq5-10$ per cent \citep{steidel+18,pahl+21,begley+22}, and provide clear evidence that higher {\Lya} equivalent widths, lower stellar masses and dust content, and fainter UV magnitudes all likely indicate higher $f_\mathrm{esc}^\mathrm{LyC}$ \citep{marchi+18,fletcher+19,begley+23,pahl+23}. Moreover, results from low-redshift analogues are finding success in uncovering which galaxy properties or spectral features can be used as robust indicators of non-negligible {\fesclyc} \citep[e.g., the UV spectral slope $\beta$ or properties sensitive to the neutral Hydrogen geometry, see;][]{chisholm+18,gazagnes+20,flury+22a,flury+22b,saldana-lopez22,saldana-lopez+23}.

Complimentary to studies of {\fesclyc}, significant progress has also been made establishing the ionizing photon production efficiencies of star-forming galaxies \citep[e.g., see][]{simmonds+23,simmonds+24b}. Analytic models typically require $\mathrm{log}_{10}(\xi_{\mathrm{ion}}\,/\, \mathrm{erg\,s^{-1}\,Hz})\gtrsim25.2-25.3$ for reionization to be complete by $z\sim5-6$, which is generally in agreement with inferences of $\xi_\mathrm{ion}$ based on the UV spectral slope \citep[$\beta$;][]{duncan+15,castellano+23}. However, these inferences rely heavily on assumptions about the stellar population models \citep[e.g., see][]{robertson+13,robertson+15,shivaei+18,seeyave+23}. As highlighted in \citet{eldridge+17} and \citet{stanway+18}, $\xi_\mathrm{ion}$ can vary by factors of $\approx2-3$ depending on the metallicity, assumed IMF, and whether or not binary stellar evolution is factored into the models, adding significant uncertainty to {\xiion} inferences.\\

Alternatively, probing the ionization conditions of galaxies can be achieved through measuring the strong nebular emission lines powered by the intense ionizing radiation from young stellar populations \citep{tang+19,tang+21,endsley+21,endsley+24}. For example, the H$\,\alpha$ emission, when combined with measurements of the UV continuum has successfully allowed {\xiion} to be measured across a range of redshifts \citep{bouwens+16,matthee+17a,shivaei+18,maseda+20}.\\

Extreme {\oiiihb} emission has also been a considerable focus in recent years, after a number of studies highlighted that a high proportion of confirmed LyC leaking galaxies have strong {\oiii} emission and high O$32$($\equiv$[\mbox{O\,\sc{iii}}]$\lambda\lambda4949,5007$\AA/[\mbox{O\,\sc{ii}}]$\lambda\lambda3726,3729$\AA\,flux ratio) values \citep[][however see also \citealt{izotov+21}]{vanzella+16,rivera-thorsen+17,izotov+18,fletcher+19}, which have been suggested as necessary requirements for high {\fesclyc} \citep{nakajima+20}. Coupled with the high {\xiion} found in galaxies with the most extreme {\oiii} emission \citep{chevallard+18,tang+19,onodera+20}, this provides significant motivation to investigate {\oiiihb} emission across cosmic time.\\

In this work we aim to piece together the evolution of the {\oiiihb} equivalent width distribution, {\EWoiiihb}, across the redshift range $3\lesssim z\lesssim8$, using a sample of galaxies selected from VANDELS spectroscopic survey and two JWST imaging surveys. 
By studying the evolution of {\EWoiiihb} across the redshift range $z\simeq 3-8$, we aim to provide a crucial direct insight into 
the evolution of {\xiion} into the reionization epoch.  

The structure of the paper is as follows. In Section \ref{sed:data_and_sample} we outline the VANDELS and JWST-based datasets and our sample selection process. In Section \ref{sec:physical_property_measurement} we describe the use of the \textsc{bagpipes} spectral energy distribution fitting code to infer physical properties, including the {\oiiihb} emission-line characteristics. Additionally, we also discuss our measurements of the UV spectral slopes ($\beta$). Section \ref{sec:results_ew0_distn} presents our inferences of the {\EWoiiihb} distributions across our two samples and explores how these evolve with both redshift and galaxy properties. Lastly, in Section \ref{sec:xiion} we estimate the ionizing photon production rates of our samples, and discuss the observed trends with physical properties in the context of the reionization process. We draw conclusions in Section \ref{sec:conclusions}.

Throughout the paper we adopt the following cosmological parameters: $H_{\rm{0}}=70\,\rm{km\,s^{-1}Mpc^{-1}}$, $\Omega_{\rm{m}}=0.3$, $\Omega_{\rm{\Lambda}}=0.7$ and all magnitudes are quoted in the AB system \citep{oke_gunn+83}.

\section{Data and sample selection}\label{sed:data_and_sample}
To infer the ionizing properties of the galaxy population emerging during the Epoch of Reionization, we assemble a high-redshift galaxy sample selected from the PRIMER \citep{dunlop+21} and JADES \citep{eisenstein+23a} surveys, as outlined below in Section \ref{subsec:jwst_sample}. We investigate the redshift evolution of the {\oiiihb} emission-line properties of the SFG population by comparing with a sample selected at $3.2\leq z_{\mathrm{spec}}\leq3.6$ from the VANDELS survey, as discussed in Section \ref{subsec:vandels_sample}.

\subsection{JWST-selected galaxy sample at $\mathbf{6.9\leq z\leq7.6}$}\label{subsec:jwst_sample}

Within the redshift range $6.9\leq z \leq 7.6$, the {\oiiihb} emission lines (comprising the $[\mbox{O\,\sc{iii}}]\lambda\lambda4959,5007$\AA\,doublet and the {$\lambda=4861$\AA} Balmer line) pass through the JWST NIRCam/F410M filter (with the bounds defined from the $\approx10^{\mathrm{th}}-90^{\mathrm{th}}$ cumulative transmission response). Owing to the unparalleled sensitivity provided by the JADES and PRIMER imaging, it is possible to obtain individual {\EWoiiihb} measurements as low as {$\simeq100-200$\AA}.

\subsubsection{PRIMER+JADES imaging and photometric catalogues}\label{subsubsec:jwst_imaging_and_catalogues}

The PRIMER \citep{dunlop+21} JWST Cycle-1 treasury programme spans  $\approx140\,\mathrm{arcmin^2}$ in the COSMOS field and $\approx~230\,\mathrm{arcmin^2}$ in the UDS field, with $5\sigma$ depths of $28.0-29.5$ magnitudes. The imaging covers eight NIRCam filters (F090W, F115W, F150W, F200W, F277W, F356W, F410M, F444W). We process these data using PENCIL (PRIMER Enhances NIRCam Image Processing Library, Magee at al. in prep), which builds upon the standard JWST pipeline (version 1.10.2, with pmap$>$1118) with additional routines for snowball and wisp removal, 1/f noise correction and background subtraction.

Our NIRCam data is supplemented by deep optical imaging from HST/ACS in the F435W, F606W, and F814W bands, taken as part of CANDELS \citep[Cosmic Assembly Near-IR Deep Extragalactic Legacy Survey;][]{grogin+11,koekemoer+11}.

The JWST Advanced Deep Extragalactic Survey \citep[JADES;][]{eisenstein+23a} GTO programme offers similar NIRCam coverage but is approximately $\approx1.5$ magnitudes deeper ($m_{\mathrm{AB}}\simeq29.0-30.5$). Here, we use the JADES GOODS-S DR2 imaging reductions \citep[described in][]{rieke+23}, including F335M coverage, with additional HST/ACS imaging (F775W and F850LP bands) from the Hubble Legacy Fields program \citep[HLF;][and references therein]{illingworth+16,whitaker+19b}.

To construct multi-wavelength photometric catalogues, we homogenise the point-spread functions (PSFs) of each HST/ACS and JWST/NIRCam image to match the F444W imaging \citep[e.g., see][]{mcleod+24}, to minimise potential biases in our aperture photometry arising from the wavelength-dependent PSFs. We perform this step using convolution kernels constructed from empirical PSF models based on stacks of isolated, unsaturated bright stars.

Initial photometric catalogues were produced by running \textsc{source extractor} \citep{bertin+96} in dual-image mode on the homogenised imaging, with the unconvolved NIRCam/F356W image used as the detection band. Fluxes are measured using 0.5-arcsec diameter apertures, enclosing $\simeq85$ per cent of the total flux for point sources. The fluxes are then corrected to total by scaling to the \textsc{FLUX\_AUTO} \citep{kron+80} photometry given by \textsc{source extractor}\footnote{A minimum floor of $1/0.85$ is enforced, corresponding to the correction to total flux in the case of a point-source}. A final correction of 10 per cent is added to our photometry to account for additional flux not captured by the Kron aperture \citep{mcleod+24}.

To ensure robust photometric redshift (and physical property) estimates, we restrict the initial catalogue to objects with full coverage in the $N=11$ PRIMER bands and apply photometric cuts based on the $5\sigma$ local depth, which approximately corresponds to $m_\mathrm{F356W}\lesssim28.0$ for the PRIMER imaging and $m_\mathrm{F356W}\lesssim29.5$ for JADES imaging, respectively.

Photometric uncertainties are derived from local depth maps, calculated as the scaled median absolute deviation ($\sigma_{\mathrm{local}}=1.483\times\sigma_{\mathrm{MAD}}$) of the flux distribution from the nearest $150-200$ empty-sky apertures, following \citet{mcleod+21}. Lastly, given that we aim to detect and infer the emission line properties of sources at $z\gtrsim6.9$, we exclude sources detected at $\geq2\sigma$ in the bluest filters (F435W and F606W in PRIMER, in addition to F775W for JADES).

\subsubsection{Photometric redshift estimates}\label{subsubsec:jwst_photoz}
To find the best photometric redshift for each object in our catalogue, we perform seven independent photometric-redshift runs, using two separate SED-fitting codes. We use \textsc{eazy-py} for five of the photometric-redshift estimates, each with a different template set. First, we use the default \textsc{eazy} template set defined in \citet{brammer+08}, with an additional high equivalent-width emission-line model to better account for the stronger emission lines commonly found in high-redshift galaxies \citep[e.g., see][]{erb+10,roberts-borsani+24}. We also perform two \textsc{eazy-py} runs using the template sets introduced in \citet{larson+23} and \citet{hainline+23}. These templates expand upon the FSPS and standard \textsc{eazy} templates using `bluer' models from BPASS \citep{stanway+18} and FSPS \citep{foreman-mackey+2013}, respectively, which are more optimised for matching the high-redshift galaxy population. 

For the final two \textsc{eazy-py} runs we make use of the \textsc{agn\_blue\_sfhz\_13}\footnote{See \url{https://github.com/gbrammer/eazy-photoz/tree/master/templates/sfhz}} and the \textsc{Pegase} \citep{fioc+97,fioc+19} template sets to broaden the range of models used. For each, the redshift was allowed to vary across the range $0.01\leq z\leq20$ in steps of $\Delta(z)=0.01$, and no apparent galaxy magnitude prior was applied \citep[e.g., see][]{hainline+23}. 

Two additional photometric redshift estimates were made with \textsc{LePhare} \citep{arnouts+99,ilbert+06} using the BC03 \citep{bruzal-charlot+03} and PEGASE \citep{fioc+97} model libraries, both adopting the \citet{calzetti+00}
dust attenuation law with $A_\mathrm{V}=0-6$. the The BC03 template set includes exponentially declining star-formation histories, permitting $\tau=0.1-15\,$Gyr, together with metallicities between $0.2\,Z_{\odot}$ and $Z_{\odot}$. A \citet{chabrier+03} initial mass function (IMF) is assumed. 

The best-estimate photometric redshift ($z_\mathrm{phot}$) for each galaxy was taken as the median redshift ($z_\mathrm{med}$) across the seven \textsc{eazy-py} and \textsc{LePhare} runs. 
To quantify the $z_\mathrm{phot}$ quality and calibrate the photometry for any potential zero-point (ZP) offsets and/or template-set mismatches \citep[e.g., see][]{dahlen+13}, we compare to a large ($N\gtrsim5500$) sample of spectroscopic redshifts compiled from various literature sources\footnote{These include galaxies with high-quality spectroscopic redshift flags from; VANDELS \citep[][see also Section \ref{subsec:vandels_sample}]{mclure+18,garilli+21}, EXCELS \citep{carnall+24}, MOSDEF \citep{kriek+15}, UDSz \citep{bradshaw+13,mclure+13a,maltby+16}, FRESCO \citep{oesch+23}, JADES DR2 \citep[][see also references therein]{rieke+23}
and additional sources listed in \citet{kodra+23} (see their Table 6).}. We calibrate each field (PRIMER/UDS, PRIMER/COSMOS, JADES/GOODS-S) and each template set independently, finding ZP/template offsets within $\pm5-10$ per cent on average across the multi-wavelength photometry.

\begin{figure}
    \centering
    \includegraphics[width=1\columnwidth]{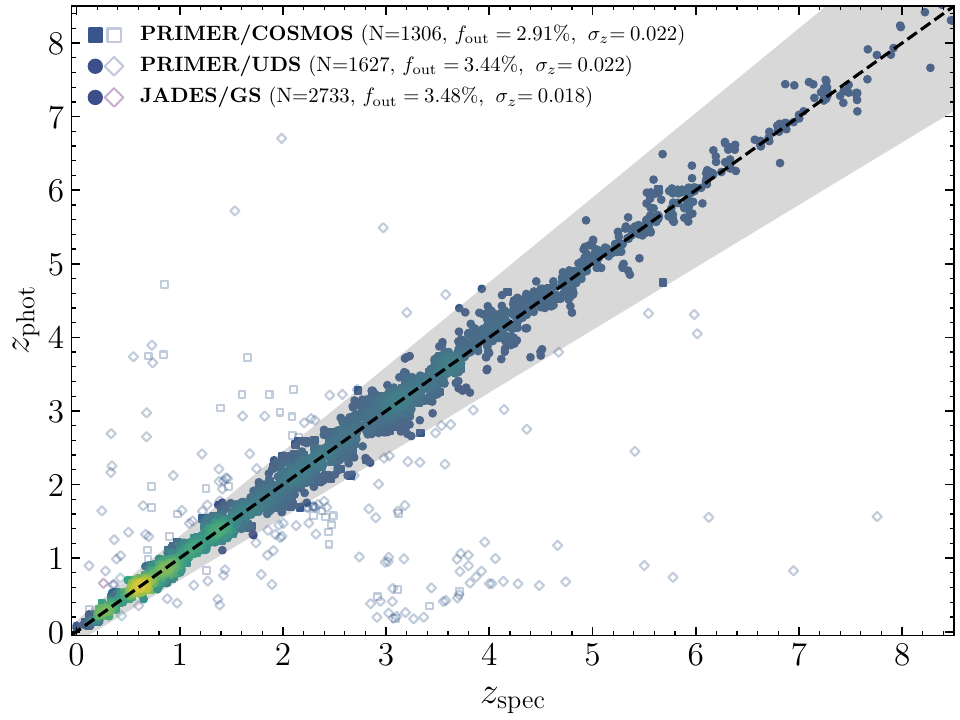}
    \caption{A comparison between photometrically estimated ($z_\mathrm{phot}\equiv z_\mathrm{med}$; see Section \ref{subsec:jwst_sample}) and spectroscopic redshifts for the subsample of $N\gtrsim5500$ galaxies with $z_\mathrm{spec}$ compiled from publicly available literature sources. For the PRIMER/COSMOS, PRIMER/UDS and JADES/GOODS-S fields, we achieve accuracies of $\sigma_z = 0.022, 0.022$ and $0.018$, and similar catastrophic outlier rates of $f_\mathrm{outlier}=2.91, 3.44$ and $3.48$ per cent, respectively. The relevant numbers of high-quality spectroscopic redshift sources are listed in the legend and the markers are coloured by source density to aid visual clarity. Hollow markers correspond to catastrophic outliers.}
    \label{fig:jwst_zphot_vs_zspec}
\end{figure}

We quantify our photometric redshift performance by $\sigma_z =1.483\times\mathrm{MAD}(\mathrm{d}z)$ (where $\mathrm{d}z=(z_\mathrm{phot}-z_\mathrm{spec})/(1+z_\mathrm{spec})$ and $\mathrm{MAD}$ is the median absolute deviation), in addition to the catastrophic outlier fraction $f_{\mathrm{outlier}}$ (where catastrophic outliers are classes as sources with $|\mathrm{d}z|\geq0.15$). Generally, across our three photometric catalogues we achieve accuracies of $\sigma_z \simeq 0.018-0.022$ and catastrophic outlier rates of $f_\mathrm{outlier}\simeq2.9-3.5$ per cent, demonstrating that our photometric redshift estimates are competitive with, or improve upon, the most robust extragalactic source redshift catalogues to date \citep[e.g., see][]{merlin+21,kodra+23,rieke+23,merlin+24,wang+24}. A comparison of our final photometric redshift estimates against the compilation of publicly available spectroscopic redshifts, in addition to the individual field $\sigma_z$ and $f_\mathrm{outlier}$ statistics, is shown in Fig. \ref{fig:jwst_zphot_vs_zspec}.
Where applicable, throughout this work we adopt a fiducial photometric redshift error of $\sigma(z_{\mathrm{phot}})\simeq 2\cdot\sigma_{z} \simeq 0.05$.

\subsubsection{Final sample of $z_{\mathrm{phot}}=6.9-7.6$ star-forming galaxies}\label{subsubsec:final_sample}
From our robust photometric redshift catalogues, we select galaxies in the range $6.9\leq z_{\mathrm{phot}}\leq 7.6$, within which the JWST/F410M band is sensitive to {\oiiihb} emission. We then visually inspect this sample of galaxies in each of the HST and JWST imaging bands (both convolved and unconvolved imaging), in addition to their best-fitting \textsc{bagpipes} SED models (see Section \ref{sec:physical_property_measurement}). We remove any sources with photometry contaminated by bright nearby objects, noise spikes, or other artefacts that may cause spurious photometric measurements and thus impact the posterior SED model. 

Moreover, we remove galaxies ($N\leq20$) that could be classified as "little red dots" (LRDs) through their extreme red colours \citep[e.g., see][]{kocevski+23}. This is motivated by the highly uncertain photometric redshifts typically seen in LRDs and the challenging nature of accurately measuring their physical properties. 
Following our initial photometric criteria outlined in Section \ref{subsubsec:jwst_imaging_and_catalogues}, in addition to the imaging and \textsc{bagpipes} posterior SED model visual inspections, we select $N=234$ galaxies from the two PRIMER fields, and $N=233$ from JADES/GOODS-S.

Lastly, comparing our sample to measurements of the UV luminosity function at $z\sim7$ \citep[e.g., see][]{finkelstein+15,bouwens+21}, we find that we are $\gtrsim90\%$ photometrically complete at UV magnitudes $M_{\mathrm{UV}}\leq-19.25$ for PRIMER and $M_{\mathrm{UV}}\leq-18.0$ for JADES. We therefore limit our analysis to objects satisfying these UV magnitude criteria, and select a final JWST-based sample of $N=279$ galaxies, with $N=132$ from PRIMER and $N=147$ from JADES.

\subsection{A VANDELS galaxy sample at $\mathbf{3.2\leq z_{\mathrm{spec}}\leq3.6}$}\label{subsec:vandels_sample}

A baseline comparison sample of star-forming galaxies outside of the reionization epoch is constructed from the VANDELS survey. Specifically, the ground-based $K-$band photometry at $\lambda_\mathrm{obs}\simeq2.2\,\mathrm{\mu m}$ available in VANDELS is sensitive to {\oiiihb} emission at $3.2~\leq~z~\leq~3.6$.

\subsubsection{The VANDELS survey}\label{subsubsec:vandels_survey}
The VANDELS spectroscopic survey is a large ESO public programme \citep{mclure+18,pentericci+18} using the \textit{VLT}/VIMOS spectrograph. In total, $N\approx2000$ spectroscopic redshifts for galaxies in the range $1.0\leq\ z_\mathrm{spec}\leq7.0$ were measured using $R\approx600$, ultra-deep optical spectroscopy spanning $0.48\,\mu\mathrm{m}\leq\lambda_\mathrm{obs}\leq 1.0\,\mu\mathrm{m}$.

VANDELS targets originate from four catalogues across two fields: the CDFS (Chandra Deep Field South) and UDS (UKIDSS Ultra Deep Survey) fields, each with a HST-based catalogue in their central region together with a wider area, ground-based photometry catalogue \citep{mclure+18}. For the HST-selected targets we use the available multi-wavelength photometry catalogues described in \citet{guo+13} (CDFS) and \citet{galametz+13} (UDS), which are based on the CANDELS programme \citep{grogin+11,koekemoer+11}. For the ground-based sources we adopt the photometry in the final VANDELS public data release \citep[DR4, described in][]{garilli+21}.

The VANDELS sample also benefits from \textit{Spitzer}/IRAC imaging in the $3.6\,\mathrm{\mu m}$ and $4.5\,\mathrm{\mu m}$ channels \citep[originally taken as part of the Spitzer Extended Deep survey;][]{ashby+13} across all four of the catalogues, providing an important anchor at wavelengths redward of {\oiiihb}, ensuring robust continuum flux estimates. Overall, all VANDELS galaxies benefit from deep, multi-wavelength photometry spanning $0.35\,\mu\mathrm{m}\lesssim\lambda_\mathrm{obs}\lesssim 4.5\,\mu\mathrm{m}$.

\subsubsection{Final sample of $z_\mathrm{spec}=3.2-3.6$ star-forming galaxies}
For this work, we select star-forming galaxies in the redshift range $3.2\leq z_\mathrm{spec}\leq 3.6$, within which the {\oiiihb} emission lines are contained within the $\simeq90$ per cent transmission regions for each of relevant $K-$band filter profiles  (Hawk-I/$K_\mathrm{S}$ for the HST-based catalogues, VISTA/$K_\mathrm{S}$ for CDFS-GROUND and WFCAM/$K$ for UDS-GROUND).

We also impose the additional requirement of having a $z_\mathrm{flag}=3,4$ or 9 redshift quality flag to select only galaxies with robust spectroscopic redshifts, having a $\gtrsim95$ per cent probability of being correct \citep[e.g., see][]{garilli+21}. Lastly, to ensure we are able to place robust constraints on the emission-line properties and obtain accurate \textsc{bagpipes} model SED posteriors (e.g., see Cochrane et al. in prep), we require all galaxies in our sample to satisfy the following photometric cuts; a $\geq3\sigma$ detection in the relevant $K-$band, and $\gtrsim3\sigma$ in at least one of the IRAC/3.6$\mathrm{\mu m}$ or IRAC/4.5$\mathrm{\mu m}$ flux measurements\footnote{We note that $\gtrsim87\,$per cent of the VANDELS sample satisfy a more conservative $K_\mathrm{S}\geq5\sigma$ cut, however the final results presented in this work are not impacted (see Section \ref{sec:results_ew0_distn}), with the typical {\EWoiiihb} consistent at the $\Delta\sim0.02\,$dex level.}.

Using the spectroscopic and photometric criteria outlined above, we obtain a final sample of $N=253$ star-forming galaxies at redshifts $3.2\leq z_{\mathrm{spec}}\leq3.6$, with robust spectroscopic redshifts and associated $N\geq11$ band photometry.

\vspace{-0.5cm}
\section{Physical properties with \textsc{bagpipes}}\label{sec:physical_property_measurement}
In this section we outline the fiducial \textsc{bagpipes} model-fitting procedure that we use to estimate each galaxy's best-fitting posterior SED model and physical characteristics (see Section \ref{subsec:bagpipes_fits}), including the photometrically inferred {\EWoiiihb} equivalent widths (see Section \ref{subsec:ewoiiihb}). We also detail the method used to measure the UV continuum slope $(\beta_\mathrm{UV})$, in Section \ref{subsec:uv_continuum_slope}.

\begin{figure}
    \centering
    \includegraphics[width=\columnwidth]{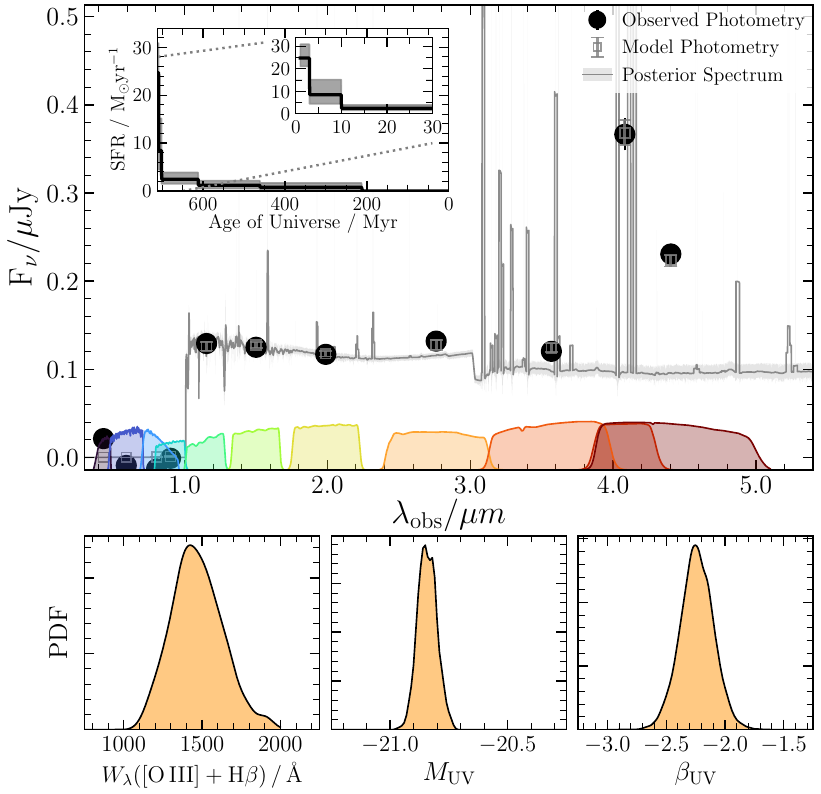}
    \caption{Spectral energy distribution for UDS-27409 fit using \textsc{bagpipes}, a $z\simeq7.3$ galaxy in our high-redshift JWST-selected sample from PRIMER/UDS. The inset in the upper panel shows the inferred star-formation history that best-fits this galaxy assuming a non-parametric continuity SFH (e.g., see Section \ref{sec:physical_property_measurement}). The measured {\oiiihb} equivalent width (\EWoiiihb), absolute UV magnitude ($M_{\mathrm{UV}}$) and UV continuum slope ($\beta_\mathrm{UV}$) posterior distributions for UDS-27409 are shown in the bottom panels. With {\EWoiiihb $\,\simeq1500\,$\AA}, this galaxy is classified as one of the extreme emission-line galaxies (EELGs) in our sample (e.g., see Section \ref{sec:results_ew0_distn}).}
    \label{fig:sed}
\end{figure}

\subsection{SED fitting with \textsc{bagpipes}}\label{subsec:bagpipes_fits}

To self-consistently measure each galaxy's physical properties (e.g., stellar mass, dust attenuation and absolute UV magnitude) alongside their nebular emission line properties (e.g., \EWoiiihb) we fit their multi-wavelength photometry using the Bayesian SED modelling code \textsc{bagpipes} \citep[Bayesian Analysis of Galaxies for Physical Inference and Parameter EStimations, see][for a full discussion of the \textsc{bagpipes} code implementation]{carnall+18,carnall+19}.

For both the VANDELS and PRIMER$+$JADES samples used in this work, we test a number of permutations of \textsc{bagpipes} model fitting configurations, varying the assumed stellar population synthesis (SPS) templates, star-formation histories (SFHs) and dust attenuation law prescriptions. A detailed comparison of how each configuration performs is presented in Appendix \ref{appendix:bagpipes_assumptions}, whereas we focus here on the fiducial \textsc{bagpipes} setups adopted for the subsequent analysis. For each sample we adopt the SPS, SFH, and dust attenuation configuration that best recovers the true, spectroscopically measured, {\EWoiiihb} (e.g., see Section \ref{subsec:spec_measurements}).

\begin{table*}
\caption{Free parameters of the fiducial \textsc{Bagpipes} models fitted to the available photometry for:\textbf{(a)} the VANDELS sample, and \textbf{(b)} the PRIMER$+$JADES sample.}
\begingroup
\setlength{\tabcolsep}{6pt}
\renewcommand{\arraystretch}{1.1}
\begin{tabular}{llllllll}
\textbf{(a) VANDELS}\\
\hline
Component & Parameter & Symbol / Unit & Range & Prior & \multicolumn{2}{l}{Hyper-parameters} \\
\hline
General & Redshift & $z$ & Fixed at $z_\mathrm{spec}$ & & & & \\
\hline
SFH & Total stellar mass formed & $M_*\ /\ \mathrm{M_\odot}$ & ($10^{\,6}$, $10^{12}$) & Logarithmic & & \\
(Delayed$-\tau$) & Stellar metallicity & $Z_*\ /\ \mathrm{Z_\odot}$ & (0.005, 1) & Uniform & & \\
& Timescale & $\tau_{\mathrm{delayed}} / \mathrm{Gyr}$ & (0.01, 15) & Uniform \\
& Age & $t_{\mathrm{Universe}}(z) / \mathrm{Gyr}$ & (0.01, $t_{\mathrm{obs}}$) & &  $t_{\mathrm{obs}}=t_{\mathrm{age}}(z)$\vspace{2mm} \\
\hline
Dust & $V-$band attenuation & $A_\mathrm{V}$ / mag & (0, 3) & Uniform & & \\
\hline
Nebular & ionization parameter & $\mathrm{log}(U)$ & ($-4$, $-1$) & Uniform & & \\
\hline
\hline
\\
\textbf{(b) PRIMER+JADES}\\
\hline
Component & Parameter & Symbol / Unit & Range & Prior & \multicolumn{2}{l}{Hyper-parameters} \\
\hline
General & Redshift & $z$ & ($z_\mathrm{phot}-3\sigma$, $z_\mathrm{phot}+3\sigma$) & Gaussian & $\mu=z_\mathrm{phot}$ $\:$ $\sigma=0.05$ \\
\hline
SFH & Total stellar mass formed & $M_*\ /\ \mathrm{M_\odot}$ & ($10^{\,6}$, $10^{12}$) & Logarithmic & & \\
(Continuity) & Stellar metallicity & $Z_*\ /\ \mathrm{Z_\odot}$ & (0.005, 1) & Uniform & & \\
& SFR change ($i\rightarrow i+1$) & $\Delta_{i}(\mathrm{log}_{10}(\mathrm{SFR}))$ & ($-$10, 10) & Student$-t$ & Default as in \citet{leja+19} \\
\hline
Dust & $V-$band attenuation & $A_\mathrm{V}$ / mag & (0, 3) & Uniform & & \\
\hline
Nebular & Ionization parameter & $\mathrm{log}(U)$ & ($-4$, $-1$) & Uniform & & \\
\hline
\hline
\end{tabular}
\endgroup
\label{table:bagpipes_params}
\end{table*}

\subsubsection{\textsc{bagpipes} fitting of the VANDELS sample}\label{subsubsec:bagpipes_fits_vandels}
In Section \ref{subsubsec:spec_measurements_nirv}, we describe the sample of ($N=29$) galaxies with spectroscopic {\EWoiiihb} measurements from NIRVANDELS \citep[e.g., see][]{cullen+21,stanton+24} that we use to calibrate the photometric {\EWoiiihb} inferences from our VANDELS galaxies.

For this VANDELS sample, the \textsc{bagpipes} configuration that best recovers the spectroscopic {\EWoiiihb} (e.g., see Fig. \ref{fig:ew0_calibration}) uses the BPASS v2.2.1 SPS models \citep{eldridge+17,eldridge+22}, with the default BPASS initial mass function (IMF) and metallicities spanning the range $5\times 10^{-3}\leq Z_{\odot} \leq 1$. The SFH is modelled as a delayed$-\tau$ model ($\mathrm{SFR(t)}\sim t\cdot e^{-t/\tau_{\mathrm{delay}}}$), parameterised by $\tau_\mathrm{delay}$ with a flat prior range between $0.01-15\,$Gyr, and $t_{\mathrm{age}}$ between $0.01$ Gyr $-\ t_\mathrm{Universe}(z)$. Nebular emission is included in the model and parameterised by the ionization parameter $\mathrm{log}(U)$, which we allow to vary between $-4$ and $-1$. A Calzetti dust attenuation law is used to account for the impact of dust \citep{calzetti+00}, and we permit $V-$band attenuation values in the range $0 \leq A_\mathrm{V} \leq 3$. The redshift is fixed to the spectroscopic redshift throughout.

\subsubsection{\textsc{bagpipes} fitting of the PRIMER$+$JADES sample}\label{subsubsec:bagpipes_fits_jwst}
We find that the \textsc{bagpipes} configuration adopted for the VANDELS sample also performs reasonably well in recovering the spectroscopic {\EWoiiihb} values of the higher-redshift PRIMER$+$JADES sample (here, we direct the reader to Section \ref{subsubsec:bagpipes_fits_jwst} for an overview of the $N=25$ galaxies for which we have spectroscopic measurements of {\EWoiiihb} from JWST/NIRSpec).
However, with the aim of maximising our ability to infer accurate {\oiiihb} equivalent widths (e.g., see Fig. \ref{fig:ew0_calibration}) we instead use a configuration adopting a non-parametric `continuity' SFH \citep[e.g., see][]{leja+19} with the BPASS SPS models. In the implementation of the continuity SFH, we use five time bins; $0-3\,$Myr, $3-10\,$Myr, $10-100\,$Myr, $100-250\,$Myr and $250-500\,$Myr, which is broadly similar to existing literature studies \citep[e.g., see][]{whitler+23,endsley+24}. We note from \citet{leja+19}, that the inferred SFH is largely insensitive to the number of time bins used in this prescription, provided that $N_\mathrm{bin}\geq4$.

It is worth emphasising that the inclusion of a finer bin spacing at more-recent look-back times (e.g., $0-3\,$Myr, $3-10\,$Myr), compared with the original youngest bin defined in \citet{leja+19} ($0-10\,$Myr), is physically motivated by the recent observational evidence for burstier SFHs in the high-redshift galaxy population \citep{looser+23,strait+23}, and that the population being observed are preferentially viewed in a `bursting phase' \citep{sun+23}. Moreover, given that strong {\oiiihb} emission traces ongoing ($\lesssim5\,\mathrm{Myr}$) star formation, shorter time bins in the most recent phase of a galaxy's SFH allows \textsc{bagpipes} to more robustly explore posterior solutions in which nebular emission contributes more strongly to the overall SED shape.

For our high-redshift galaxy sample, we find the BPASS models recover the spectroscopic {\EWoiiihb} measurements with marginally more scatter; however opting for BPASS over the fiducial BC03 SPS model choice has no significant impact on our analysis. For the BPASS configuration, we use the associated \textsc{cloudy} \citep{ferland+13} nebular emission models in which $\mathrm{log}(U)$ can vary in the range $-4\leq \mathrm{log}(U)\leq -1$.

Given the photometric selection of the JWST sample, we fit each galaxy with a Gaussian redshift prior centered on the best estimate $z_\mathrm{phot}$, with error $\sigma(z_\mathrm{phot})\simeq0.05$, as described in Section \ref{subsubsec:jwst_photoz}. This allows any physical parameter estimates to fold in the impact of the photometric redshift uncertainty.

Lastly, we also use a Calzetti dust law in the optimal PRIMER$+$JADES \textsc{bagpipes} configuration, with the same $A_\mathrm{V}$ range.
A more thorough discussion of the impact of the adopted dust attenuation law, as well as other configuration choices, can be found in Appendix \ref{appendix:bagpipes_assumptions}. A summary of the full fiducial \textsc{bagpipes} model parameters and their priors for the two samples is provided in Table. \ref{table:bagpipes_params}, and an example best-fitting posterior SED model is shown in Fig. \ref{fig:sed}.

\begin{figure*}
    \centering
    \includegraphics[width=2.1\columnwidth]{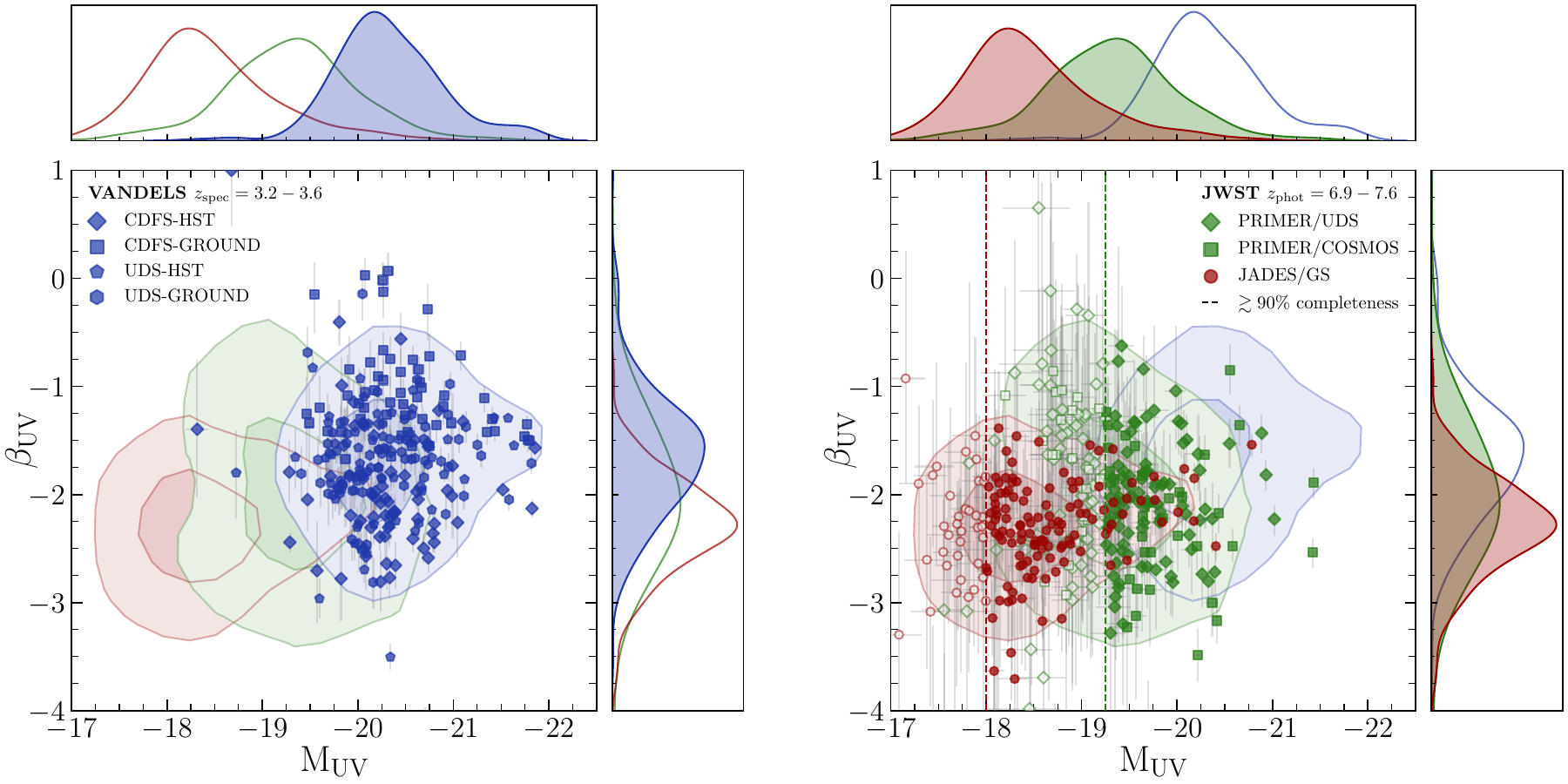}
    \caption{\textbf{Left:} Absolute UV magnitude ($M_\mathrm{UV}$) and UV continuum slope ($\beta_\mathrm{UV}$) for the $z_\mathrm{spec}=3.2-3.6$ galaxy sample selected from the VANDELS spectroscopic survey (blue markers). The source photometric catalogue for each VANDELS source is indicated by the marker shape (see legend) and contours from the right panel are shown to highlight the overlap in parameter space. \textbf{Right:} Same as left panel, but for the JWST sample at $z_\mathrm{phot}=6.9-7.6$ selected from the PRIMER/COSMOS and PRIMER/UDS fields (green markers), and the JADES/GOODS-S field (red markers). Contours representing the $68^\mathrm{th}$ and $95^\mathrm{th}$ percentile boundaries are shown for each sample on both panels, in their respective colours, to better illustrate the overlap in parameter space between the intermediate- and high-redshift samples. We also show the 1D marginal probability density distributions of $M_\mathrm{UV}$ and $\beta_\mathrm{UV}$ for each sample in the sub-panels of each figure. The 90\% completeness limits in $M_\mathrm{UV}$ from our photometrically selected PRIMER ($M_\mathrm{UV}\leq-19.25$) and JADES ($M_\mathrm{UV}\leq-18$) samples are marked with vertical dashed lines.}
    \label{fig:sample_muv_vs_beta}
\end{figure*}

To calculate the absolute rest-frame UV magnitude ($M_{\mathrm{UV}}\equiv M_{1500}$) for each object, we first draw 1000 model SEDs from their \textsc{bagpipes} fit posteriors and then integrate each model through a {$\Delta\lambda=100\,$\AA} top-hat centered on $\lambda_{\mathrm{rest}}=1500\,$\AA\:  \citep{donnan+22,begley+23}. Each galaxy's $M_{\mathrm{UV}}$ is then given as the median of their posterior-derived $M_{\mathrm{UV}}$ distribution, with $\pm1\sigma$ errors quoted as the $16^\mathrm{th}$ and $84^\mathrm{th}$ percentiles.
The $M_\mathrm{UV}$ distributions of the two samples are plotted alongside their UV continuum slope estimates ($\beta_\mathrm{UV}$; see Section \ref{subsec:uv_continuum_slope}), in Fig. \ref{fig:sample_muv_vs_beta}. 

It is clear from Fig. \ref{fig:sample_muv_vs_beta} that the PRIMER dataset used in this work is crucial in providing the important overlap in $M_\mathrm{UV}$ between the intermediate- and high-redshift samples needed to disentangle the $M_\mathrm{UV}$ and redshift dependencies (e.g., see Section \ref{subsec:distribution_with_properties}). On the other hand, the use of both PRIMER and JADES provides the broad dynamic range that is paramount for a full exploration any $M_\mathrm{UV}$ dependence (e.g., see Section \ref{subsubsec:uvmag_dependence} and Section \ref{sec:xiion}).  

\subsection{The UV continuum slope}\label{subsec:uv_continuum_slope}

The UV continuum slope, $\beta$, where $f_{\lambda}\propto\lambda^\beta$, offers key insights into the properties of the high-redshift galaxy population. As highlighted by  \citet{cullen+23}, blue UV slopes ($\beta\lesssim -2.2$) are indicative of galaxies that are young, relatively low metallicity, with low levels of dust attenuation \citep[or are even dust free, e.g., see][]{cullen+24}. Such blue UV slopes indicate that these galaxies likely have higher-than-average ionizing photon production efficiencies \citep{cullen+24,topping+24}, and moreover, are thought to indicate non-negligible ionizing photon escape fraction \citep{begley+22,chisholm+22,choustikov+24,kreilgaard+24}. It is therefore  vital to link emission line properties, including {\EWoiiihb}, to the UV continuum slope, if we are to better understand the main galaxies contributing to reionization.
\\ 

The UV continuum slopes of the galaxies in our PRIMER sample are measured  following the method outlined in \citet{cullen+24}, briefly described here. Firstly, we take the galaxy photometry probing rest-frame wavelengths {$\lambda_{\mathrm{rest}}\leq 3000\,$\AA} (i.e., selecting filters with {$\lambda_{95^{\mathrm{th}}}\leq3000\,$\AA}, where  $\lambda_{95^{\mathrm{th}}}$ is the $95^{\mathrm{th}}$ percentile of the cumulative filter transmission curve). We then model this photometry as a power law ($f_{\lambda}=\alpha\cdot\lambda^{\beta}$), with full IGM attenuation adopted below {$\lambda_{\mathrm{rest}}<1216\,$\AA} given our sample is at $z\gtrsim7$. In addition, the possible effect of any $\mathrm{Ly}\alpha$ damping wing present is modelled using equation 2 of \citet{miralda+00}. In total, this model has four free parameters: (i) the UV continuum slope, $\beta$; (ii) the flux normalisation factor of the power law, $\alpha$; (iii) the redshift of the galaxy, $z$; and (iv) the neutral hydrogen fraction of the surrounding IGM, $x_{\mathrm{HI}}$. We note here that $x_{\mathrm{HI}}$ (spanning values of $0\leq x_{\mathrm{HI}}\leq1$), is effectively a dummy parameter that we marginalise over as photometric data alone has little-to-no constraining power on this parameter. Readers are directed to \citet{cullen+24} for additional discussion of the UV slope measurement methodology.

To sample the posterior distributions of the parameters we use Monte Carlo (MCMC) ensemble sampler \textsc{emcee} \citep{foreman-mackey+2013}, adopting a Gaussian prior on the redshift, $\mathcal{N}(z,\sigma)$, where $z=z_{\mathrm{phot}}$, the best-estimate photometric redshift and $\sigma=\sigma(z_{\mathrm{phot}})$ determined in Section \ref{subsubsec:jwst_photoz}, respectively. We note that the typical $\beta$ measurements are unchanged when fixing at the best-estimate $z_{\mathrm{phot}}$, with the semi-flexible $z$ parameter prescription allowing photometric redshift errors to propagate through to $\beta$. The UV continuum slope is allowed to vary in the range $-10\leq\beta\leq 10$, with a uniform prior adopted.

For VANDELS we follow an almost identical method, with the exception that we replace the complete IGM attenuation (and $\mathrm{Ly}\alpha$ damping parameterisation) with the IGM prescription detailed in \citet{inoue+14}. As the \citet{inoue+14} IGM attenuation prescription is that of the \textit{average} transmission function at a given redshift $z$, in practice we implement this model addition with two parameters ($z$,$\phi$),  where $\phi$ is a scaling factor accounting for the stochasticity of IGM transmission at $z\sim3-4$ \citep[e.g., see][]{inoue+14,begley+23}, which we marginalise over. As with the high-redshift model fitted to our PRIMER sample, removing this parameter does not significantly impact our $\beta$ measurements.

\subsubsection{Typical physical properties of our galaxy samples}\label{subsubsec:typical_physical_properties}
As shown in Fig. \ref{fig:sample_muv_vs_beta}, the median absolute UV magnitude of our full JWST-selected sample at $z_\mathrm{phot}=6.9-7.6$ is $\langle M_\mathrm{UV}\rangle=-19.3$, with the $2.5^\mathrm{th}-97.5^\mathrm{th}$ percentile ($\pm2\sigma$) range being $M_\mathrm{UV}=[-18.1,-20.4]$. Within the JWST-selected sample, we also find a $\Delta(M_\mathrm{UV})\simeq1.2$ magnitude difference in the typical $M_\mathrm{UV}$ between the PRIMER and JADES subsamples (as expected given the relative imaging areas and depths of the surveys). Specifically, the PRIMER sample has $\langle M_\mathrm{UV}\rangle=-19.6$ (and $M_\mathrm{UV}=[-19.3,-20.9]$) and the JADES sample has $\langle M_\mathrm{UV}\rangle=-18.5$ (and $M_\mathrm{UV}=[-18.0,-20.1]$).

We measure an average UV continuum slope for the JWST sample of $\langle\beta_\mathrm{UV}\rangle=-2.19$ and a $P_{2.5-97.5}(\beta_\mathrm{UV})=[-1.33,-3.02]$. We additionally recover a very mild $\mathrm{d}\beta/\mathrm{d} M_\mathrm{UV} \simeq -0.17$ evolution between the samples ($\langle\beta_\mathrm{UV}\rangle=-2.08$ for PRIMER; $\langle\beta_\mathrm{UV}\rangle=-2.28$ for JADES), which is consistent with recent measurements of the UV continuum slopes of $z\gtrsim8$ galaxies \citep[e.g., see][]{cullen+24,topping+24}. \\

The $z\simeq3.2-3.6$ sample from VANDELS has a median absolute UV magnitude of $\langle M_\mathrm{UV}\rangle=-20.26$ with a ($2.5^\mathrm{th}-97.5^\mathrm{th}$ percentile) range of $M_\mathrm{UV}=[-19.56,-21.42]$. The median UV continuum slope is $\langle\beta_\mathrm{UV}\rangle=-1.64$, with a $\beta_{\mathrm{UV}}=[-0.73,-2.56]$ $\pm2\sigma$ range. This value is moderately redder than the JWST sample at $z\sim6.9-7.6$ ($\Delta(\beta)\simeq0.6$) as expected given the $z-\beta_\mathrm{UV}$ evolution previously observed over $4\lesssim z\lesssim9$ \citep{rogers+13,bouwens+15,cullen+23}.

From our SED fits, we find our samples span $\approx3\,$dex in stellar mass, which motivates exploring the mass-dependence of {\EWoiiihb} in Section \ref{sec:results_ew0_distn}. Our JWST sample has a typical inferred stellar mass of $\langle\mathrm{log}_{10}(M_*/\mathrm{M}_{\odot})\rangle\simeq 8.3$, with the PRIMER and JADES subsamples having typical stellar masses $\simeq0.3\,$dex higher and lower respectively. The range of masses probed in these high-redshift samples is $\mathrm{log}_{10}(M_*/\mathrm{M}_{\odot})\approx 7.4-9.3$.  On the other hand, we find that galaxies in our VANDELS sample at $z=3.2-3.6$ have inferred stellar masses in the range $\mathrm{log}_{10}(M_*/\mathrm{M}_{\odot})\approx 8.7-10.1$, with a median of $\langle\mathrm{log}_{10}(M_*/\mathrm{M}_{\odot})\rangle\simeq 9.3$. We highlight that the stellar mass values quoted above are all from the best-estimate delayed$-\tau$ SFH model \textsc{bagpipes} fits for consistent comparison between samples. The inferred stellar mass is sensitive to the assumed SFH prescription, with the equivalent continuity SFH \textsc{bagpipes} fits implying stellar masses $\approx 0.2-0.3\,$dex higher in the JWST sample, and $\approx 0.05-0.1\,$dex higher in the VANDELS sample.

\subsection{{\oiiihb} equivalent width measurements}\label{subsec:ewoiiihb}
The aim of fitting each galaxy with \textsc{bagpipes}, after careful consideration of the most optimal model configuration, was to self-consistently infer the {\oiiihb} emission-line and physical properties from the available multi-wavelength photometry. To generate {\EWoiiihb} posterior distributions for each galaxy we first draw $\sim 10^3$ model SEDs from the resulting \textsc{bagpipes} posteriors. For each SED model instance, we measure the continuum flux following the method  outlined in Section \ref{subsubsec:spec_measurements_jwst} to make our spectroscopic {\EWoiiihb} measurements. The associated {\oiiihb} line fluxes are then directly pulled from the \textsc{bagpipes} model posteriors, with {\EWoiiihb} given as described in Section \ref{subsubsec:spec_measurements_jwst}. 
Lastly, we apply corrections derived from the best-fitting spectroscopic-photometric {\EWoiiihb} comparisons presented in Section \ref{subsec:spec_measurements}, including propagation of the associated uncertainties.

Below, in Section \ref{sec:results_ew0_distn}, we discuss the sample statistics of the inferred {\EWoiiihb} values, and explore how the measured {\EWoiiihb} distributions vary as a function of physical properties.

\subsection{A sample of spectroscopic {\oiiihb} measurements}\label{subsec:spec_measurements}
To ensure our photometric {\oiiihb} equivalent width inferences are accurate, we validate our measurements using a subsample of galaxies that have rest-frame optical spectroscopy available, for both our VANDELS and PRIMER+JADES samples.

\subsubsection{The NIRVANDELS sample at $z_\mathrm{spec}=3.2-3.6$}\label{subsubsec:spec_measurements_nirv}

A subset of the full VANDELS spectroscopic sample benefits from near-IR spectroscopy covering {\oiiihb} from NIRVANDELS. This survey is a near-IR follow-up programme using Keck/MOSFIRE \citep{mclean+12} and VLT/KMOS \citep{davies+13,sharples+13}, targeting $N\approx70$ VANDELS galaxies at $3.0~\lesssim~z_\mathrm{spec}~\lesssim~3.8$ with $H{-}\,/\,K-$band spectroscopy spanning $1.5\,{\mu\mathrm{m}}\lesssim\lambda_\mathrm{obs}\lesssim 2.4\,{\mu\mathrm{m}}$. The reader is directed to \citet{cullen+21} and \citet{stanton+24} for a full description of the target selection, data reduction procedures and spectroscopic line measurements of the NIRVANDELS MOSFIRE and KMOS datasets, respectively. 

In total, the additional NIR observations allow direct spectroscopic measurements of both the [\mbox{O\,\sc{iii}}] and $\mathrm{H\,\beta}$ lines for $N=65$ galaxies in the $2.95 \leq z_\mathrm{spec} \leq 3.8$ redshift range, of which $N=29$ are also selected in the $3.2 \leq z_\mathrm{spec}\leq 3.6$ VANDELS sample used here. 

\begin{figure}
    \centering
    \includegraphics[width=1\columnwidth]{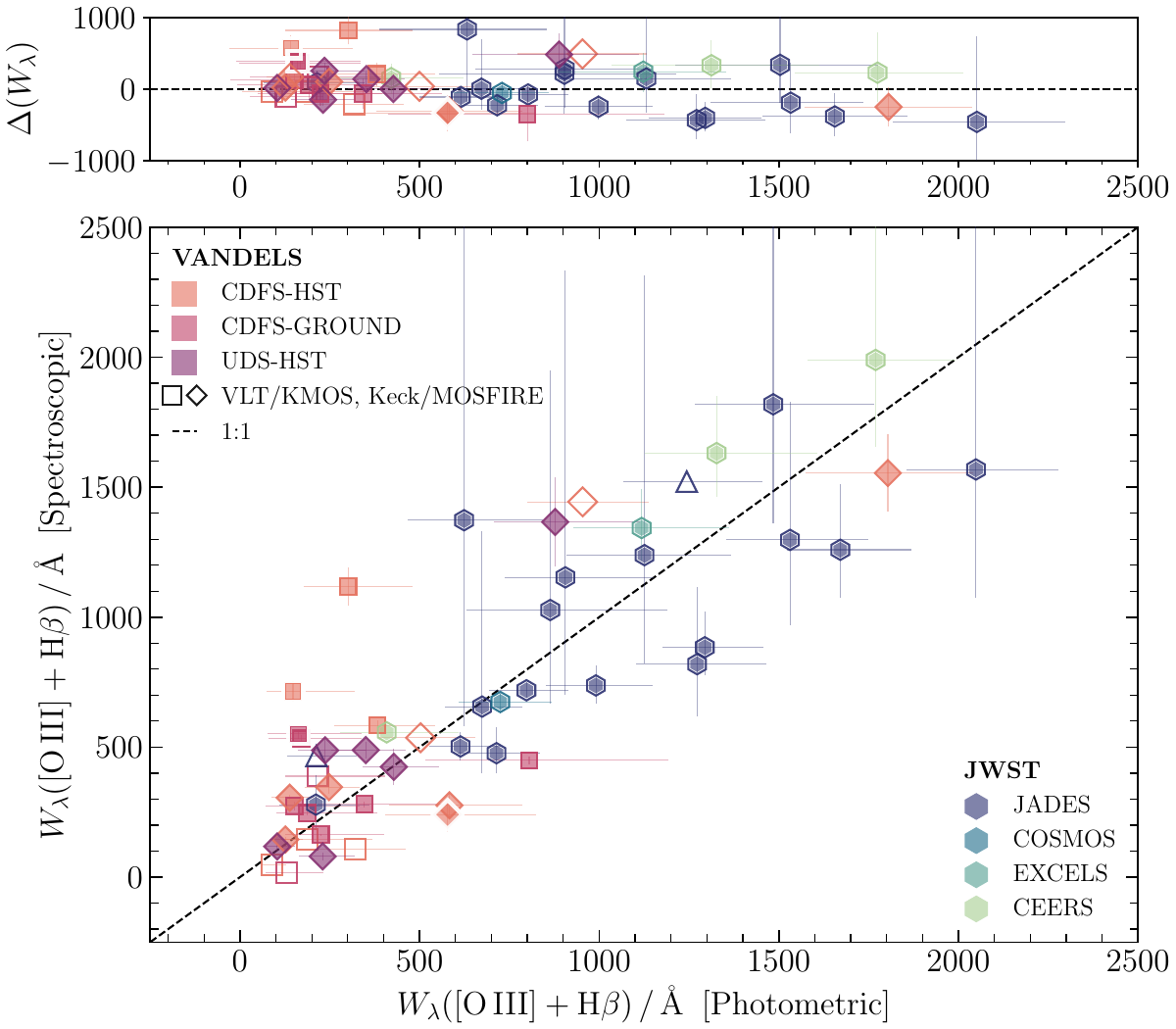}
    \caption{A comparison of the photometrically inferred {\EWoiiihb} measurements from our fiducial \textsc{bagpipes} SED models (e.g., see Section \ref{subsec:ewoiiihb}) against spectroscopic measurements for the $N=29$ galaxies from NIRVANDELS (Section \ref{subsubsec:spec_measurements_nirv}) and the $N=21$ PRIMER$+$JADES galaxies with available spectroscopic {\EWoiiihb} measurements, in addition to two further literature objects (see Section \ref{subsubsec:spec_measurements_jwst}). We find that our {\EWoiiihb} measurements are in excellent agreement with those from spectroscopy across a wide dynamic range of {\EWoiiihb} ($\approx500-2000\,$\AA\,in the JWST sample and $\approx50-1000\,$\AA\, in the VANDELS sample). For VANDELS, the markers are coloured by their VANDELS photometric catalogue (red-purple), with NIRVANDELS Keck/MOSFIRE and VLT/KMOS spectropic measurements denoted by diamonds and squares respectively. Empty markers represent spectroscopic {\EWoiiihb} upper limits, whilst white outlines highlight galaxies that do not satisfy the additional $K-$band and IRAC $3.6\mu\mathrm{m}\:/\:4.5\mu\mathrm{m}$ photometric criteria. For the high-redshift sample, markers are coloured by their spectroscopy origins, and empty triangles are spectroscopic {\EWoiiihb} upper limits.}
    \label{fig:ew0_calibration}
\end{figure}

A comparison between the direct spectroscopic {\oiiihb} equivalent width values and our photometrically inferred {\EWoiiihb} measurements is shown in Fig. \ref{fig:ew0_calibration}. Although only containing a relatively small subset of the total sample, we find excellent agreement between the photometrically inferred {\EWoiiihb} measurements from our fiducial \textsc{bagpipes} model and those measured from NIRVANDELS spectroscopy. The median offset ($\Delta(W_{\lambda})=W_{\lambda,\mathrm{spec}}-W_{\lambda,\mathrm{phot}}$) in the sample is {$\simeq90\pm100\,$\AA}, with the fitted linear relation used to correct our photometric {\EWoiiihb} measurements given as; $W_{\lambda,\mathrm{spec}} = (0.94\pm0.16)\times W_{\lambda,\mathrm{phot}} + (86\pm84$\,\AA).

\subsubsection{Literature JWST spectroscopy at $z=6.9-7.6$}\label{subsubsec:spec_measurements_jwst}
Having established the robustness of our VANDELS sample {\EWoiiihb} measurements, in this section we outline the spectroscopic subset used to validate the {\EWoiiihb} measurements of our JWST-selected PRIMER+JADES sample. 

From the total sample of $N=279$ galaxies selected from the PRIMER and JADES programmes in the redshift range $6.9~\leq~z~\leq~7.6$, a subset of $N=21$ galaxies have publicly available JWST NIRSpec/MSA spectroscopy capable of providing sufficient {\EWoiiihb} constraints. The majority of this subsample ($N=18$) have spectra released as part of JADES DR3 \citep{deugenio+24}. We opt to use the available prism spectroscopy over the higher resolution grating spectroscopy, with the goal of obtaining more accurate constraints of the underlying continuum flux and thus more robust {\EWoiiihb} measurements. Where applicable, the measured line fluxes are compared to those from the higher-resolution NIRSpec data and found to be fully consistent (as expected given that strong rest-frame optical lines will be less impacted by lower resolution spectroscopy).

In addition to the galaxies with JADES DR3 spectroscopy, we include spectroscopic measurements for a PRIMER/COSMOS target queried from the public DAWN JWST Archive (DJA) repository\footnote{See \url{https://dawn-cph.github.io/dja/} for access to the repository. PRIMER/COSMOS target credit; PI: D. Coulter, DD6585. Three additional spectra credit: PI: S. Finkelstein, ERS1345 and PI: P. Arrabal Haro, DD2750.} and a PRIMER/UDS medium resolution grating spectrum from EXCELS \citep[][Scholte et al. in prep]{carnall+24}. Lastly, to increase the size of the spectroscopic validation sample, we include two further prism spectra queried from the DJA.

To measure {\EWoiiihb} from the prism spectra, we first fit a power-law continuum from {$4825$\AA$\,\leq \lambda_{\mathrm{rest}}\leq5050\,$\AA}, excluding the emission line regions. Next, we subtract this continuum and fit each emission line with a Gaussian profile using \textsc{specutil}. If H$\beta$ or [\mbox{O\textsc{iii}}]$\lambda4959$ are not significantly detected, we fix their respective wavelengths, and, in the latter case, also constrain the line to the theoretical line ratio ($\simeq2.98$). 

The final equivalent widths are the sum of individual line widths, calculated as {$(1+z_\mathrm{spec})^{-1} \cdot (f_{\mathrm{line}}\,/\,f_\mathrm{cont})_i$}, where $f_\mathrm{line}$ and $f_\mathrm{cont}$ are from the Gaussian and continuum fits, respectively.
Lastly, we visually inspect each spectrum to ensure each galaxy has reasonable fits to both the line profiles and underlying continuum.

We show the comparison between our spectroscopic {\EWoiiihb} measurements and those from our fiducial \textsc{bagpipes} photometric inferences in Fig. \ref{fig:ew0_calibration}. Although the $z\gtrsim 7$ spectroscopic sample features few galaxies with {\EWoiiihb\,$\lesssim$\,500\,\AA}, we have a wide dynamic range including numerous EELGs ({\EWoiiihb\ $>1000\,$\AA}) and find a small median offset of {$\Delta(W_\lambda)\simeq-90\pm100\,$\AA}. Again, we 
fit a linear relation between the spectroscopic and photometric measurements; $W_{\lambda,\mathrm{spec}} = (0.92\pm0.12)\times W_{\lambda,\mathrm{phot}} - (33\pm102$\,\AA), which we apply as a correction to our {\EWoiiihb} measurements.\\

Overall, across both the NIRVANDELS and JWST-based spectroscopic subsamples, we find that we recover accurate {\EWoiiihb} values from our fiducial \textsc{bagpipes} photometric model fits. Importantly, we robustly distinguish between populations of SFGs with extreme {\oiiihb} emission lines ({\EWoiiihb\ $>1000\,$\AA}) and those with weaker equivalent widths ({\EWoiiihb\,$\lesssim200\,$\AA}).

\section{The {\oiiihb} equivalent width distribution }\label{sec:results_ew0_distn}

\begin{figure}
    \centering
    \includegraphics[width=1\columnwidth]{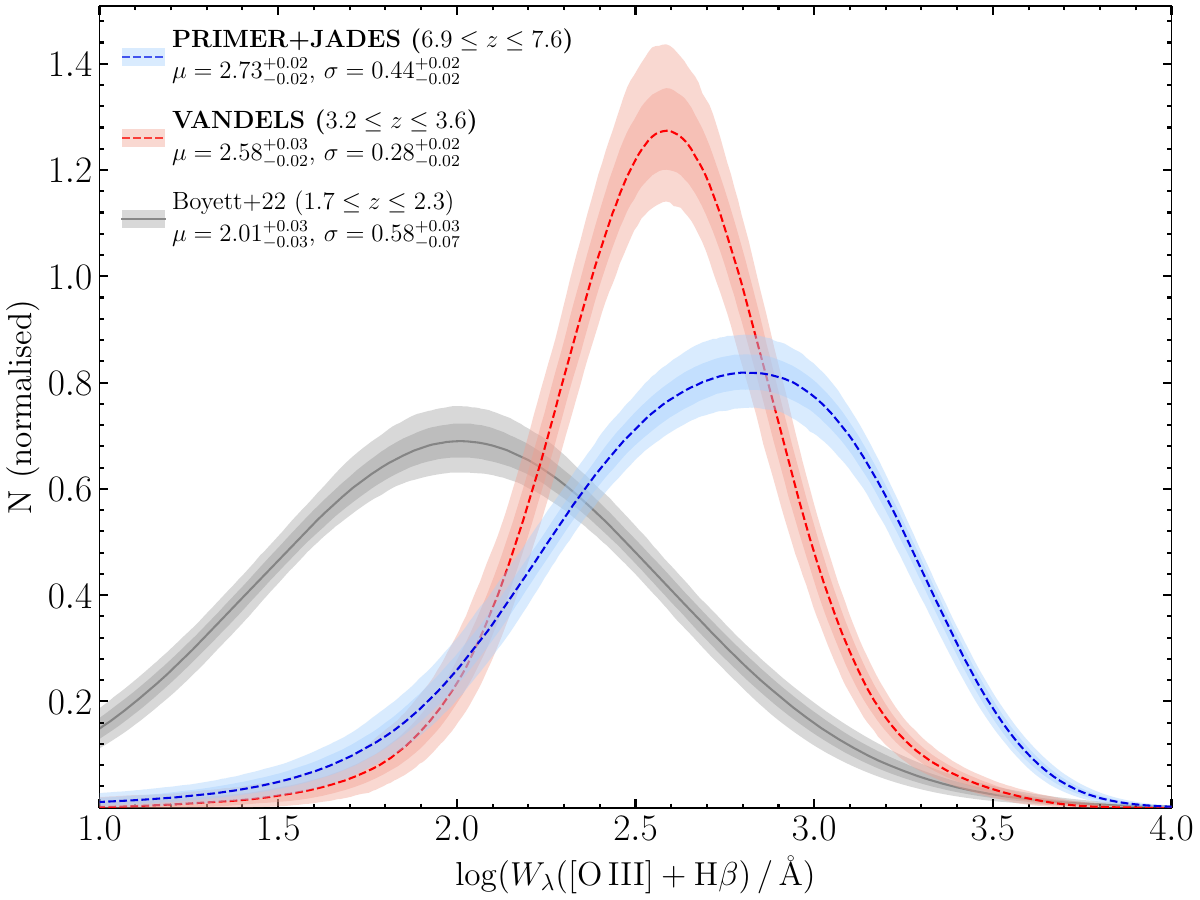}
    \caption{The {\oiiihb} equivalent-width distributions for our $3.2~\leq~z_\mathrm{spec}~\leq~3.6$ VANDELS sample (red) and our photometric, JWST-selected sample from PRIMER and JADES at $6.9\leq z_{\mathrm{phot}}\leq7.6$ (blue). The distributions are shown as kernel density estimates, computed from Monte-Carlo sampling the corrected {\EWoiiihb} posterior distributions from the best-fitting \textsc{bagpipes} SED models. The distribution medians are shown with dashed lines, and the $\pm1\sigma$ and $\pm2\sigma$ confidence intervals are shown as darker and lighter shaded regions. The sample median {\EWoiiihb} values and distribution widths (in $\mathrm{log}_{10}-$space) are listed in the legend. To reference results at lower redshifts, we also show the $z\simeq2$ {\EWoiiihb} distribution from \citet{boyett+22}.}
    \label{fig:ew0_distn_main}
\end{figure}

In Fig. \ref{fig:ew0_distn_main}, we show the {\oiiihb} equivalent width distributions for the full JWST ($\gtrsim90\%$ photometrically complete, $N=279$) and VANDELS-selected ($N=253$) samples. The {\EWoiiihb} equivalent width distributions are presented in the form of kernel density estimates: Monte-Carlo sampling from the individual {\EWoiiihb} posterior distributions produced from our fiducial \textsc{bagpipes} fits (see Section \ref{subsec:bagpipes_fits}) and their respective spectro-photometric calibrations derived in Sections \ref{subsubsec:spec_measurements_nirv} and \ref{subsubsec:spec_measurements_jwst}. The dashed lines denote the median KDE, whilst the darker and lighter shaded regions highlight the $\pm1\sigma$ and $\pm2\sigma$ KDE confidence regions, respectively.

The KDE representations of the {\EWoiiihb} distributions better show their true underlying shape, and highlight that our high-redshift sample follows a {\EWoiiihb} distribution that deviates from the typically observed log-normal form. 

The median {\oiiihb} equivalent width of the full JWST sample (shown in the legend of Fig. \ref{fig:ew0_distn_main} in $\mathrm{log}_{10}-$space for easier comparison with the existing literature) is $\mu\equiv\mathrm{log}_{10}($\EWoiiihb/\AA$)=2.73^{+0.02}_{-0.02}$ ({\EWoiiihb} $\simeq 540\pm25\,$\AA), with the $\pm1\sigma$ confidence limits here and throughout Section \ref{sec:results_ew0_distn} being estimated from our Monte-Carlo sampling. The width of the {\EWoiiihb} distribution is calculated as the scaled median absolute deviation (again, for each Monte-Carlo realisation), from which we find $\sigma\equiv\sigma(\mathrm{log}_{10}($\EWoiiihb/\AA$))=0.44\pm0.02$ (implying a $\pm1\sigma$ range of $P({\pm1\sigma})\approx195-1480\,$\AA). Given the relatively asymmetry seen in the high-redshift {\EWoiiihb} distribution (see Section \ref{subsec:asymmetric_distn}), we highlight that the modal value is \EWoiiihb$\simeq660\,${\AA}.

For the VANDELS spectroscopic sample at $z=3.2-3.6$, we measure a {\EWoiiihb} distribution with $\mu=2.58\pm0.02$ ({\EWoiiihb}$\simeq 380\pm18\,$\AA), and $\sigma=0.28\pm0.02$ (implying $P({\pm1\sigma})\approx200-700\,$\AA). 

\subsection{Literature results at $\mathbf{z\simeq2}$}\label{subsec:literature_z2}
To assess the evolution of the {\EWoiiihb} distribution across a broader dynamic range in redshift, we compare with the $z\simeq2$ [\mbox{O\,\sc{iii}}]$\lambda5007$\AA\ equivalent width distribution measured by \citet{boyett+22} for a magnitude-limited ($M_\mathrm{UV}\leq-19.0$) sample of $N=672$ galaxies with rest-frame optical spectroscopy \citep[see][]{brammer+12,oesch+18b}.
Fitting a log-normal functional form to the \textit{observed} $W_{\lambda}$([\mbox{O\,\sc{iii}}]) distribution, \citet{boyett+22} find best-fitting parameters of $\mu_\mathrm{LN}=1.84\pm0.03$ and $\sigma_\mathrm{LN}=0.58\pm0.03$ (quoted in base 10 log-space).\\

To directly compare the \citet{boyett+22} results with our measurements, we must first convert their $W_{\lambda}$([\mbox{O\,\sc{iii}}]) distribution to a {\EWoiiihb} distribution. To account for the [\mbox{O\,\sc{iii}}]$\lambda4960$\AA\ line flux, we adopt the theoretical line ratio [\mbox{O\,\sc{iii}}]$\lambda5007$\AA\,/\,[\mbox{O\,\sc{iii}}]$\lambda4960$\AA$\simeq 2.98$ \citep[e.g.,][]{storey+00}. 
Establishing the contribution from the $\mathrm{H\,\beta}$ emission line is less straightforward due to observational evidence suggesting the [\mbox{O\,\sc{iii}}]$\lambda5007$\AA\,/\,$\mathrm{H\,\beta}$ ratio evolves with redshift and stellar-mass \citep{kewley+15,cullen+16,dickey+16}. 
In this analysis we opt to use the empirically measured $W_{\lambda}(\mathrm{H\,\beta})\simeq0.115\times W_{\lambda}$([\mbox{O\,\sc{iii}}]$\lambda5007$\AA$)^{1.065}$ relation presented in \citet{boyett+22} (based on the results of \citealt{tang+19}), giving a conversion factor $f_\mathrm{conv}\simeq1.48$. This shifts the location parameter in the best-fitting log-normal distribution by $\Delta(\mu_\mathrm{LN})\simeq0.17\,$dex to $\mu_\mathrm{LN}\simeq2.01$ (i.e., corresponding to a typical {\oiiihb} equivalent width of $\simeq100\,${\AA}).

\subsection{Evolution of the {\EWoiiihb} distribution}
Contrasting the VANDELS and JWST distributions, we see a clear $\simeq1.5\times$ factor increase in the median equivalent width from $z\simeq3.4$ to $z\simeq7.3$, in addition to a strong increase in the {\EWoiiihb} distribution width of $\simeq0.16\,$dex. This evolution in the average {\EWoiiihb} across our sample is broadly expected, as galaxies become increasingly younger and more metal-poor towards higher-redshifts \citep[e.g.,][]{cullen+19,langerooddi+22}. 

Comparing our VANDELS sample to the {\EWoiiihb} distribution inferred by \citet{boyett+22} from a spectroscopic sample of $z\simeq2$ SFGs, we see a {$\mathrm{d}$\EWoiiihb$/\mathrm{d}z\simeq200\,$\AA} evolution with redshift. This compares to a shallower evolution of {$\mathrm{d}$\EWoiiihb$/\mathrm{d}z\simeq40\,$\AA} moving from the VANDELS to PRIMER$+$JADES samples (noting that this is approximately equivalent to a power-law in time with $\alpha\sim-0.5$). We further discuss the origin of the apparent redshift evolution, including any impact on an underlying $M_\mathrm{UV}$-dependence in Section \ref{subsec:distribution_with_properties}.

Across the $z\approx2-8$ redshift range shown in Fig. \ref{fig:ew0_distn_main}, we also see a clear change in the observed width of the {\EWoiiihb} distributions. Broadly, we observe that the {\EWoiiihb} distribution width sharply narrows by $\sim0.3\,$dex between $z\simeq2-3.5$ ($\sigma=0.58^{+0.03}_{-0.07}$ to $\sigma=0.28\pm0.02$), before moderately broadening again by $\sim0.16\,$dex to $\sigma=0.44\pm0.02$ at $z\simeq7$. Together with the evolution in the median {\EWoiiihb}, these results suggest that the population of high-redshift, lower-mass galaxies is dominated by stochastic SFHs \citep[and potentially increased $f_{\mathrm{esc}}(\mathrm{LyC})$ and/or decreased metallicity;][]{boyett+24,endsley+24}. This population gives way to more evolved populations with higher stellar masses and more-typical star-formation histories at $z\sim3.5$.

\subsection{The emergence of an asymmetric {\EWoiiihb} distribution}\label{subsec:asymmetric_distn}
Aside from the strong shift in the typical {\EWoiiihb} of galaxies with redshift, the most stark dissimilarity between the lower-redshift ($z\simeq2-4$) and $z\gtrsim6.5$ {\oiiihb} equivalent width distributions is the divergence from a log-normal-like distribution shape (e.g., clearly visible in Fig. \ref{fig:ew0_distn_main}).

\begin{figure}
    \centering
    \includegraphics[width=\columnwidth]{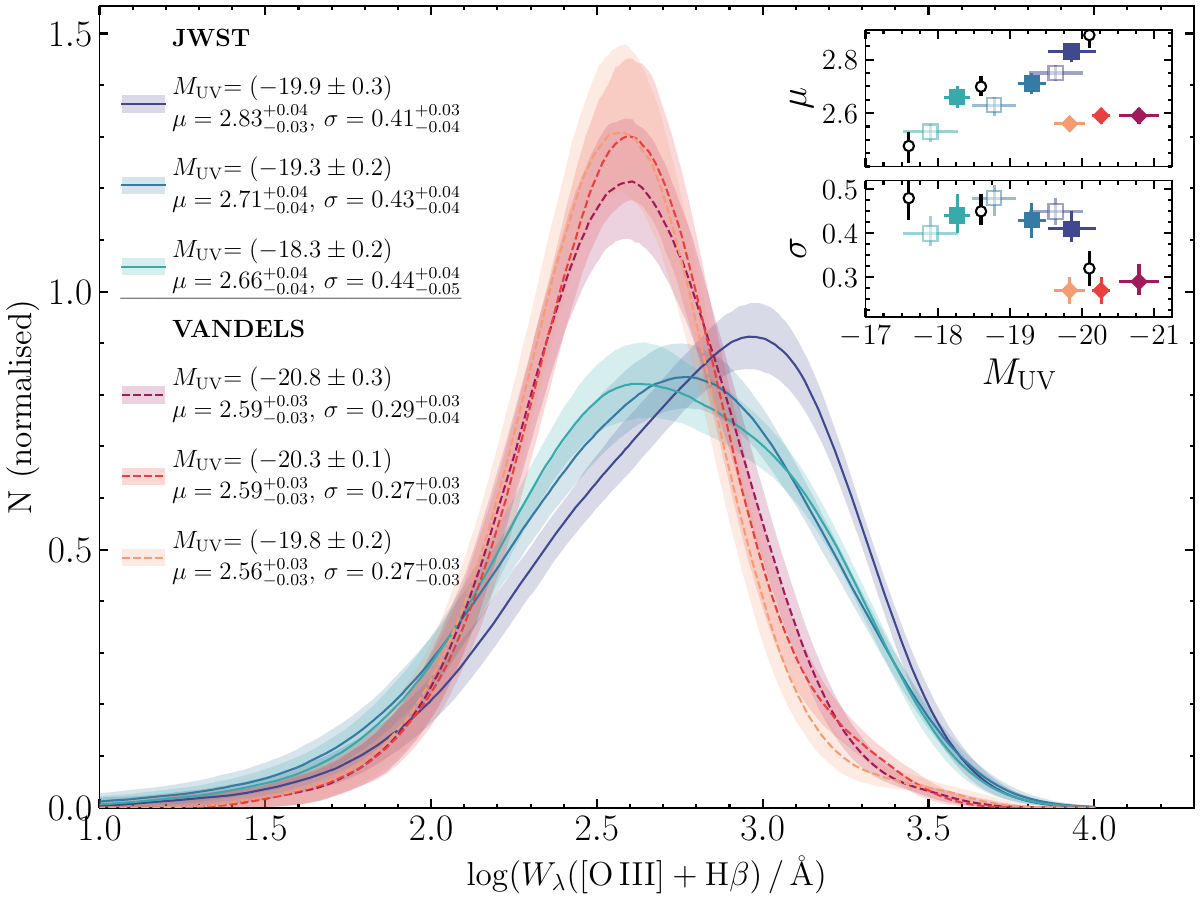}
    \includegraphics[width=\columnwidth]{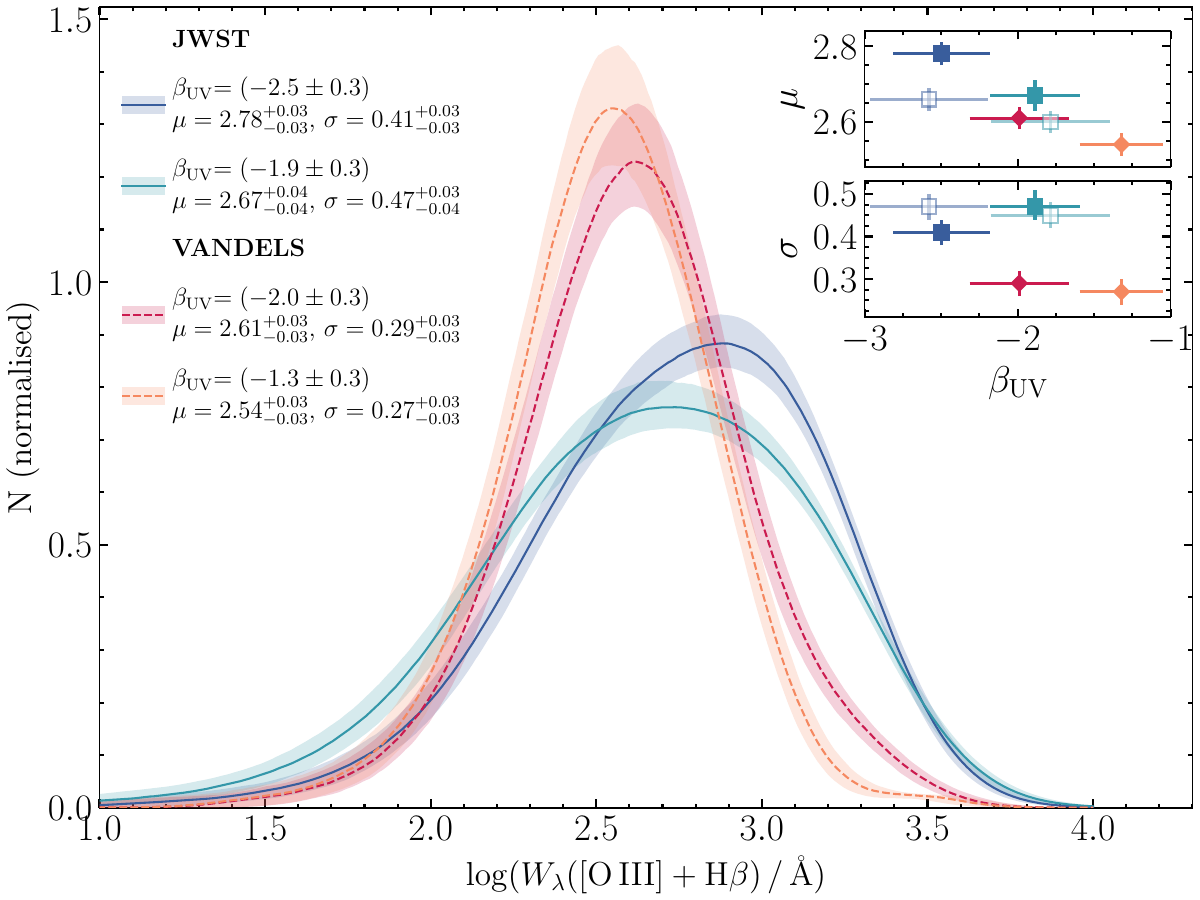}
    \includegraphics[width=\columnwidth]{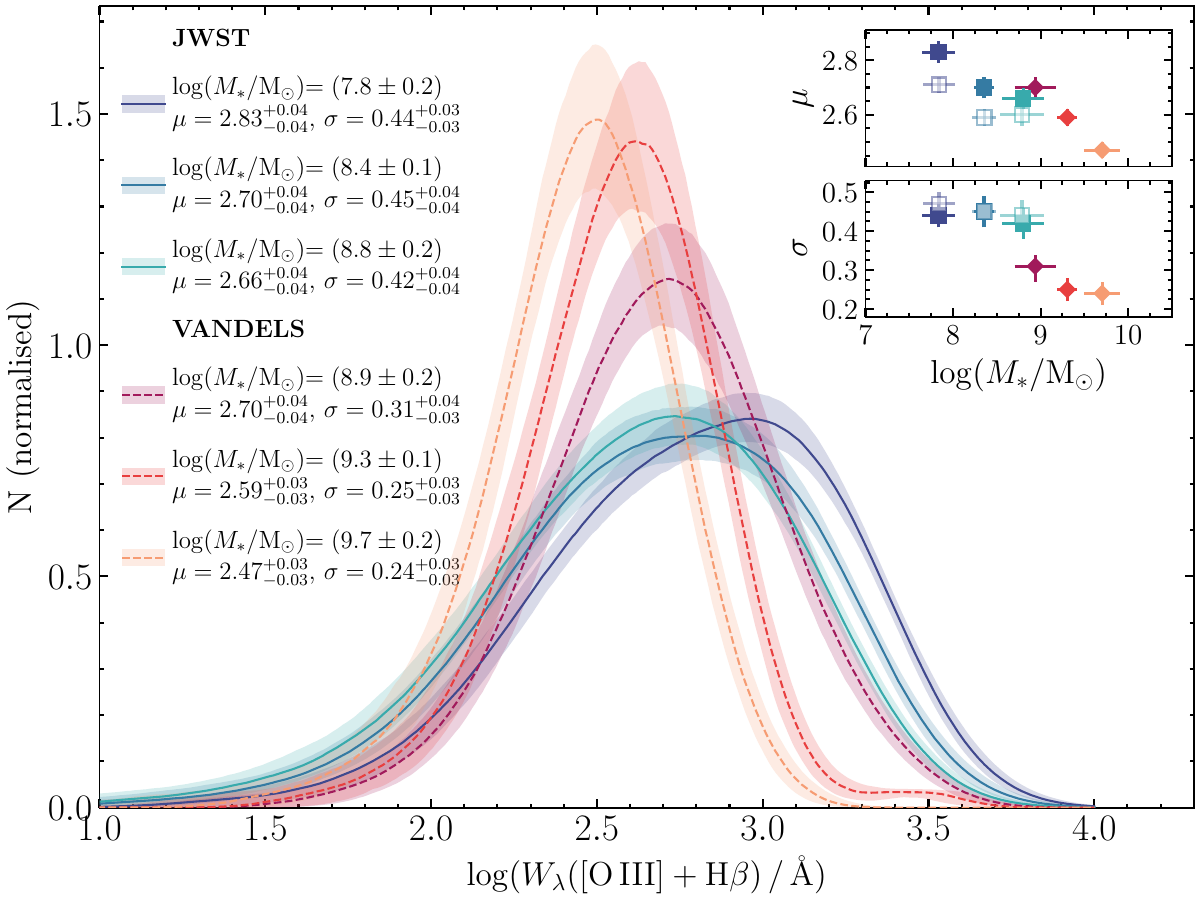}
    \vspace{-0.5cm}\caption{The {\oiiihb} equivalent-width distributions for the VANDELS (yellow-red colour scheme) and JWST-selected (blue-purple colour scheme) galaxies, split into equally sized subsamples based on their absolute UV magnitudes ($M_\mathrm{UV}$; \textbf{top panel}), UV continuum slopes ($\beta_\mathrm{UV}$; \textbf{centre panel}) and inferred stellar-masses ($\mathrm{log}(M_{*}/\mathrm{M_\odot})$; \textbf{bottom panel}). The physical property median and $\sigma_\mathrm{MAD}$ for each subsample split and the associated {\EWoiiihb} distribution median ($\mu$) and width ($\sigma$) is listed in the legend of each panel. The upper and lower inset figures of each panel illustrate the evolution in $\mu$ and $\sigma$, respectively, with non-filled markers indicating the values when not limiting our sample to $90\%$ photometric completeness. In the inset figures of the top panel, we also plot the results of \citet{endsley+24} with black circles. Overall, we find low-mass SFGs that are relatively blue and UV-bright produce the strongest {\oiiihb} emission at high redshift.}
    \label{fig:ew0_distn_physicalparametersplit}
\end{figure}

As expected in the presence of a true underlying log-normal distribution, for the VANDELS sample we find $\simeq51.4^{+2.9}_{-2.3}$ per cent of the sample falls below the peak ($\mu\simeq2.58$) of the {\EWoiiihb} distribution. For the high-redshift PRIMER$+$JADES sample however, we see a marked deviation from a log-normal distribution, with $\simeq58.4^{+3.0}_{-4.0}$ per cent of the sample falling below the distribution peak (i.e., below the mode of the distribution; $\mu\simeq2.82$). This asymmetric tail towards lower {\EWoiiihb} is tentative, differing from the expected $\simeq50$ per cent at the $\simeq2\sigma$ level, but becomes more statistically significant ($~\simeq5\sigma$) towards brighter UV luminosities (see brightest $M_\mathrm{UV}$ subset in the top panel of Fig. \ref{fig:ew0_distn_physicalparametersplit}).

Such an increase in the relative fraction of galaxies with lower {\EWoiiihb} is consistent with recent literature evidence that star-formation at higher redshifts is burstier \citep{looser+23,strait+23,sun+23,faisst+24,endsley+24,simmonds+24}, with a greater proportion of the population likely to be observed in a `down-phase' of star formation.

\subsection{Dependence of the {\EWoiiihb} distribution on physical properties}\label{subsec:distribution_with_properties}

The exact nature of the star-forming galaxy populations that dominated the ionizing photon budget required to drive reionization remains an open question \citep[e.g., see][]{robertson+23}. Namely, the relative contributions of the total photon budget from populations across the the UV luminosity function, as well as the amount of these photons that can subsequently escape, is still hotly debated in the literature \citep{matthee+22,naidu+22b,prieto-lyon+23}.

Motivated by the clear signal that strong {\oiiihb} emission is indicative of a strongly ionizing environment \citep{tang+19,boyett+24,simmonds+24}, here we investigate the dependence of the {\EWoiiihb} distribution on other physical properties of our samples to gain insight into the nature of the potential drivers of reionization.

\subsubsection{Absolute UV magnitude, $M_\mathrm{UV}$}\label{subsubsec:uvmag_dependence}
To assess the dependence of {\EWoiiihb} on {$M_\mathrm{UV}$}, we construct three equally sized subsets from our PRIMER$+$JADES galaxy sample with median $M_\mathrm{UV}$ $\pm\sigma_{\mathrm{MAD}}$  of  $\langle M_\mathrm{UV} \rangle=-19.9\pm0.3,\, -19.3\pm0.2,\, -18.3\pm0.2$, and infer their respective {\EWoiiihb} distributions, as shown in the top panel Fig. \ref{fig:ew0_distn_physicalparametersplit} (blue-purple lines). In the inset panels, we show the evolution of the {\EWoiiihb} distribution median and width as a function of $M_{\mathrm{UV}}$, highlighting the clear increase in the typical {\EWoiiihb} value, and decrease in the population {\EWoiiihb} scatter in UV-bright galaxies. Within our high-redshift sample, we see a $\simeq0.17\,$ dex evolution from $M_\mathrm{UV}\simeq-18$ to $M_\mathrm{UV}\simeq-20$, corresponding to a {$\mathrm{d}${\EWoiiihb}$/\mathrm{d}M_{\mathrm{UV}}\sim-140\,$} dependence. This is in excellent agreement with the {\EWoiiihb}$-M_\mathrm{UV}$ trend seen by \citet{endsley+24} in a sample of $z\simeq6-9$ SFGS selected from JADES (plotted as black circles on Fig. \ref{fig:ew0_distn_physicalparametersplit}).

In contrast to the increasing average {\EWoiiihb}, the width of the distribution decreases from faint to bright absolute UV magnitude. As highlighted in \citet{endsley+24}, such a scenario is physically consistent with increasingly bursty SFHs at high redshift and/or in UV-faint galaxies \citep{atek+22,chen+24}. 

Although individual SFHs are challenging to constrain robustly through photometric SED modelling alone, the \textsc{bagpipes} (continuity SFH) fits to galaxies in the fourth $M_{\mathrm{UV}}$ quartile ($M_{\mathrm{UV}}\gtrsim-18.4$) show an average rise in the most recent $3\,$Myr period of star-formation ($\Delta(\mathrm{log}_{10}(\mathrm{SFR\,/\,M_\odot yr^{-1}})\sim0.18\pm0.03$) that is $\approx1.5\times$ less strong compared with the first $M_{\mathrm{UV}}$ quartile subset ($M_{\mathrm{UV}}\lesssim-19.7$; $\Delta(\mathrm{log}_{10}(\mathrm{SFR\,/\,M_\odot yr^{-1}})\sim0.34\pm0.04$). Taken in conjunction with the metallicity-{\EWoiiihb} anti-correlation seen in our \textsc{bagpipes} fits (and a lack of galaxies with both a high {\EWoiiihb} and ultra-low metallicity), we conclude the {\EWoiiihb$-M_\mathrm{UV}$} trends seen in our high-redshift sample are consistent with bursty star-formation modes becoming more dominant in the UV-faint population.

For the VANDELS sample at $z\simeq3.2-3.6$, we find no statistically significant trend in {\EWoiiihb} with UV luminosity, in agreement with the findings of \citet{boyett+22}. More broadly, this result is in accordance with studies of the ionizing properties of $z\simeq2-5$ galaxies (e.g., {\xiion}, and other relevant tracers such as strong {\oiiihb} and $\mathrm{H}\alpha$ emission), which find weak-to-no trends with $M_\mathrm{UV}$ \citep[e.g., see][see also Section \ref{sec:xiion}]{bouwens+16,nanayakkara+20,castellano+23}. It is worth noting that the majority of the studies carried out at these intermediate redshifts to date have limited dynamic range in $M_\mathrm{UV}$, probing galaxies at $M_\mathrm{UV}\simeq-19.5$ or brighter. Testing whether or not the lack of a {\EWoiiihb$-M_{\mathrm{UV}}$} correlation continues to ultra-faint UV magnitudes ($M_\mathrm{UV}\gtrsim-18$) (e.g., \citealt{maseda+20} hints at an extremely ionizing, $M_\mathrm{UV}-$faint population of $\mathrm{Ly}\alpha$ emitters) would require analyses of larger samples of faint $M_{\rm UV}$ populations at $z\sim2-5$. 

Lastly, the faintest low-redshift and brightest high-redshift subsamples both have a median UV luminosity of $\langle M_\mathrm{UV}\rangle\sim-19.8$. Nonetheless, as evident from Fig. \ref{fig:ew0_distn_physicalparametersplit}, there is a clear distinction in the two {\EWoiiihb} distributions ($\Delta(\mu)\sim0.27\,$dex), which is strong evidence that the global redshift evolution seen here \citep[and in other literature, e.g., see][]{boyett+22,boyett+24} is not due to the differences in sample $M_\mathrm{UV}$. 

\subsubsection{UV continuum slope, $\beta_\mathrm{UV}$}
Low metallicities, young stellar populations and modest or no dust, which are properties often seen in faint, low-mass SFGs, are conducive to bluer UV continuum slopes \citep{mclure+11,bouwens+12b,dunlop+13b,calabro+22,cullen+23,topping+24}. With such conditions come stronger ionizing properties and thus an expectation of strong {\oiiihb} emission.

To evaluate any trends in our samples with {$\beta_\mathrm{UV}$}, we construct two subsets, split on the median $\beta_\mathrm{UV}$ for each of the VANDELS ($\langle\beta_\mathrm{UV}\rangle\simeq-$1.6) and PRIMER$+$JADES ($\langle\beta_\mathrm{UV}\rangle\simeq-$2.2) samples. In the high-redshift sample we find a clear trend, going from $\mu=2.67\pm0.04$ at $\beta_{\mathrm{UV}}=-1.9\pm0.03$ to $\mu=2.78\pm0.03$ at $\beta_{\mathrm{\rm UV}}=-2.5\pm0.03$. 
In the VANDELS sample we see a weaker trend, that is also marginally shifted to lower {\EWoiiihb}, with $\mu=2.61\pm0.03$ at $\beta_{\mathrm{UV}}=-2.0\pm0.03$ and $\mu=2.55\pm0.02$ at $\beta_{\mathrm{UV}}=-1.3\pm0.03$.

Another consideration is the potential contribution to the SED shape from nebular continuum, which acts to redden the UV slope \citep[e.g., see][]{cullen+23,cullen+24,katz+24}. As highlighted in \citet{topping+24}, galaxies with the most extreme {\oiiihb} emission are expected to be marginally reddened by their nebular continuum, which would then act to flatten any $\beta_\mathrm{UV}-${\EWoiiihb} correlation. In parallel, they show that their  bluest galaxies ($\beta_\mathrm{UV}\lesssim-2.8$) show signatures of weaker {\oiiihb} emission and thus less nebular continuum. However, isolating a sample of $\beta_\mathrm{UV}\lesssim-2.8$ galaxies from our JWST sample yields no significant {\EWoiiihb} distribution differences compared to the full high-redshift sample.

\subsubsection{Stellar mass, $\mathrm{log}_{10}(M_*/\mathrm{M_\odot})$}
The observed trend in the inferred {\EWoiiihb} distribution with stellar mass across both our samples is unambiguous. For the PRIMER$+$JADES sample, after splitting into three equal-sized bins of stellar mass, we see a decrease in the median {\EWoiiihb} of $\Delta(\mu)\simeq0.13\,$dex over a $\Delta(\mathrm{log}_{10}(M_*/\mathrm{M_\odot}))\sim1.0\,$dex increase in stellar-mass (going from {\EWoiiihb$\simeq675^{+65}_{-60}\,$\AA} at $\langle \mathrm{log}_{10}(M_*/\mathrm{M_\odot}) \rangle=7.8\pm0.2$ to {\EWoiiihb$\simeq460^{+45}_{-40}\,$\AA} at $\langle \mathrm{log}_{10}(M_*/\mathrm{M_\odot}) \rangle=8.8\pm0.2$).  

The impact of stellar mass on the {\EWoiiihb} distribution in our $z\sim3.2-3.6$ VANDELS sample is even more pronounced, with {\EWoiiihb} decreasing by $\Delta(\mu)\sim0.11\,$dex per $\Delta(\mathrm{log}_{10}(M_*/\mathrm{M_\odot}))\sim0.4\,$dex increase in stellar mass across the three bins ({\EWoiiihb$\simeq 500^{+50}_{-45},\, 390^{+30}_{-25},\, 295\pm20\,$\AA} for $\langle \mathrm{log}_{10}(M_*/\mathrm{M_\odot}) \rangle=8.9\pm0.2,\, 9.3\pm0.1,\, 9.7\pm0.2$, respectively).

These results are in excellent agreement with an array of literature backing up a robust anti-correlation between stellar mass and the ionizing conditions of SFGs across a wide range of redshifts \citep[e.g., see][]{de-barros+19,tang+19,atek+22,matthee+22,llerena+23,caputi+24,chen+24,llerena+24}. It is also worth highlighting that this is fully consistent with the known correlation between stellar mass and stellar iron abundances at high redshift \citep[e.g., see][]{cullen+19,kashino+22,chartab+24,stanton+24}. 

Interestingly, we also find little evidence for any redshift evolution at fixed stellar mass, as shown by the approximately continuous trend moving from the high-redshift high-mass bin to the low-redshift, low-mass bin. The lack of any significant redshift evolution (at fixed stellar mass) is in keeping with \citet{matthee+22}, who in their Fig. 8, show an approximately constant {$\mathrm{log}_{10}(M_*/\mathrm{M_\odot})-$$W_{\lambda}$([\mbox{O\,\sc{iii}}])} relation beyond cosmic noon ($z\gtrsim2$).

An important point to note is that stellar mass is likely somewhat degenerate with {\EWoiiihb} when estimating from SED modelling to photometry alone.  However, as shown by Cochrane et al. (submitted), inaccurate stellar masses are primarily driven by the inability of SED modelling codes to account for emission lines. Therefore, given the robust recovery of {\EWoiiihb} demonstrated by our photometric-to-spectroscopic calibrations (see Fig. \ref{fig:ew0_calibration}), we do not expect this to effect to significantly impact our derived stellar masses. In addition, we do not see any systematic trends with physical properties (e.g., observed luminosity) in our {\EWoiiihb} calibration checks, and thus any systematic shifts in stellar mass should impact our samples uniformly. 

As highlighted in Section \ref{subsubsec:final_sample}, due to the nature of our photometric sample selection there are potential impacts from selection biases at the lowest stellar-masses. Specifically, low-mass galaxies are likely more readily identifiable when they show stronger emission lines, which may make the apparent {$\mathrm{log_{10}(M_*/M_\odot)}-$\EWoiiihb} trend steeper. Moreover, a further potential consideration comes from the analysis of the FIRE-2 simulations by \citet{sun+23}, who suggest that only $\sim50\%$ or less of galaxies with $\mathrm{log_{10}(M_*/M_\odot)}\lesssim8$ are detectable at $z\sim7$ in JADES-like surveys. To mitigate these potential concerns we have limited our sample to a $\gtrsim90\%$ photometric completeness limit (see Section \ref{subsubsec:final_sample}), which in turn implies we only suffer significant mass incompleteness below $\mathrm{log_{10}(M_*/M_\odot)}\lesssim7.6$ \citep[e.g., see][]{pozzetti+10,mcleod+21}. Moreover, as shown in the lower panel inset in Fig. \ref{fig:ew0_distn_physicalparametersplit}, the {$\mathrm{log_{10}(M_*/M_\odot)}-$\EWoiiihb} trend is in fact shallower when we do not account for photometric incompleteness. Nonetheless, deeper surveys with larger samples of galaxies at the lowest masses will be required to confirm these trends down to stellar masses of $\mathrm{log_{10}(M_*/M_\odot)}\sim7$ and below.

\subsection{The prevalence of extreme emission line galaxies}
Galaxies with extreme emission lines (EELGs) are potentially the most ionizing star-forming galaxies in the Universe - with low metallicities, intense periods of star formation and young O/B stellar populations, they likely produce copious amounts of ionizing photons, and are therefore extremely important to consider within the Epoch of Reionization \citep{van-der-wel+11,tang+19,eldridge+22,matthee+22,rinaldi+23,endsley+24,simmonds+24}.

\begin{figure}
    \centering
    \includegraphics[width=\columnwidth]{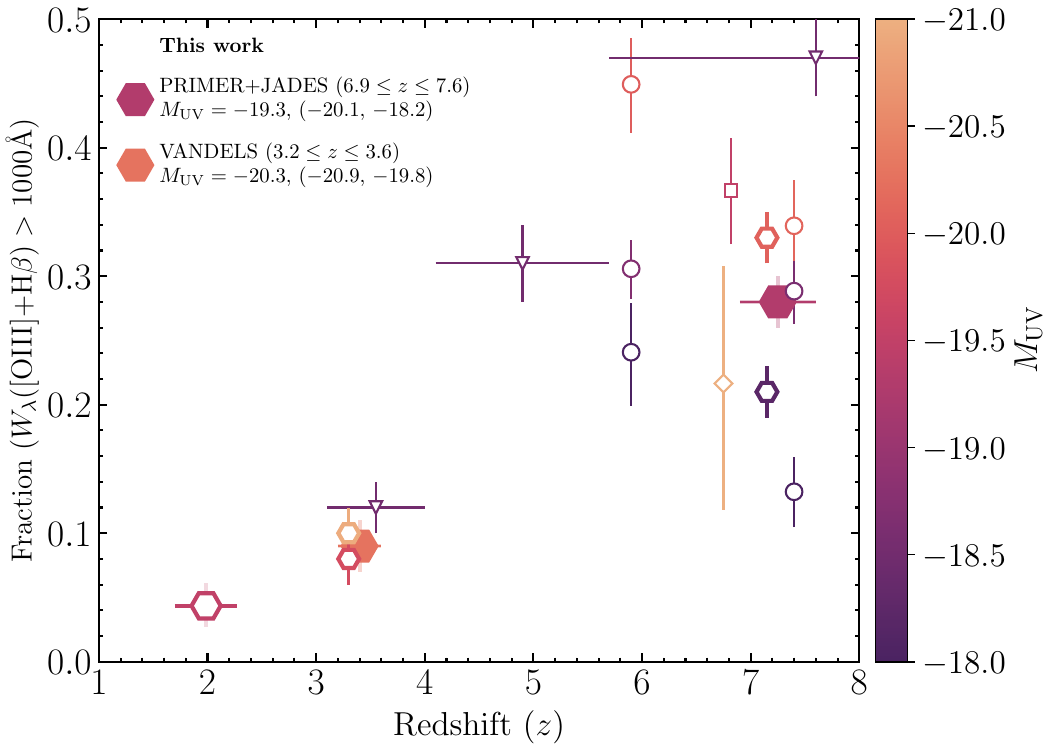}
    \caption{The extreme emission line fraction (\EWoiiihb$\ \geq1000\,$\AA) across the $2\lesssim z\lesssim 8$ redshift range. Strong redshift evolution is evident, with $f_{\mathrm{EELG}}$ increasing from $\simeq4$ per cent at $z\simeq2$ to $\simeq28$ per cent at $z\simeq7.3$. There is also a $M_\mathrm{UV}$-dependent modulation at high redshift that is weak or non-existent at low-to-intermediate redshifts. The full VANDELS and PRIMER$+$JADES samples assembled in this work are shown as large filled hexagons, while $M_\mathrm{UV}$-split subsets are shown as open hexagons. To provide a comparison to existing literature, we plot results (coloured by the sample $\langle M_\mathrm{UV} \langle$) from \citet{boyett+22} at $z\sim2$ (large, open hexagon) and from \citet{endsley+24} using JADES galaxies at $z\sim6$ and $z\sim7-9$ (open circles). Additionally, we include data from \citet{endsley+21,endsley+22} at $z\sim6.7$ (open diamond) and $z\sim6.3-8.0$ (open square), respectively. Lastly, we display recent EELG fraction measurements from \citet{boyett+24} (here treated as upper limits as they correspond to the fraction with \EWoiiihb$\ \gtrsim1135\,$\AA) as triangles.}
    \label{fig:eelg_frac_redshift_evolution}
\end{figure}

Using our measured {\EWoiiihb} distributions, we calculate the EELG fraction ($f_\mathrm{EELG}$ where {\EWoiiihb}\,$\geq1000\,${\AA}) within our VANDELS and PRIMER$+$JADES samples, as shown in Fig. \ref{fig:eelg_frac_redshift_evolution}. In our full high-redshift JWST-selected sample we find an EELG fraction of $f_\mathrm{EELG}=0.29{\pm0.02}$, with an underlying $M_\mathrm{UV}$ dependence shown by the evolution of the EELG fraction from $f_\mathrm{EELG}=0.21\pm0.02$ in our faintest bin ($\langle M_\mathrm{UV} \rangle\simeq-18.3$) to $f_\mathrm{EELG}=0.33{\pm0.02}$ in our brightest bin ($\langle M_\mathrm{UV} \rangle\simeq-19.9$). The EELG fraction, as well as the observed $M_\mathrm{UV}$ trend, are consistent with the recent JADES-based inferences by \citet{endsley+24}. Such trends with $M_{\mathrm{UV}}$ are perhaps not surprising, given that more-extreme emission lines and an enhanced UV luminosity go in tandem with the bursty SFH modes commonly seen in the high-redshift galaxy population \citep[e.g.,][]{rinaldi+23,sun+23,simmonds+24b}.

At lower redshifts there is a rapidly declining EELG fraction, reaching $f_\mathrm{EELG}=0.08\pm0.01$ for our VANDELS sample at $z\sim3.4$, a trend that continues down to $z\sim2$ and lower \citep[e.g., $f_\mathrm{EELG}\simeq4$ per cent in][]{boyett+22}. We note that this value is $\sim2\times$ less than the $\sim15$ per cent measured by \citet{llerena+23} for a sample of $N=35$ VANDELS galaxies at similar redshifts. However, that sample was selected based on strong {\mbox{C\,\sc{iii}}]$\lambda1908$\AA} emission, which generally indicates a higher ionization state, and is thus a deliberately biased subsample of the general SFG population. 

In contrast to the high-redshift sample, in the VANDELS sample we find no statistically significant ($\lesssim1\sigma$) $M_\mathrm{UV}$ dependence of the EELG fraction, with the fainter ($\langle M_\mathrm{UV} \rangle\simeq-19.8$) and brighter ($\langle M_\mathrm{UV} \rangle\simeq-20.8$) subsets being within $\pm2$ per cent of the full sample EELG fraction. The VANDELS results presented in this work play a crucial role bridging the redshift parameter space between $z\sim1-2$ and $z\gtrsim6$, and strengthen the evidence for a systematic increase in the prevalence of strong {\oiiihb} emitters from the local Universe, through cosmic noon to the reionization epoch \citep[][]{boyett+22,boyett+24,endsley+24}.

\subsection{The influence of AGN}
Active galactic nuclei (AGN) within galaxies can also drive strong {\oiiihb} emission \citep[e.g., see][]{kewley+13,coil+15}. Here, we consider the possible impact of AGN driven {\oiiihb} emission, as well as the underlying evolution in the AGN population, on the observed {\EWoiiihb} evolution between our samples. 

For the sample of galaxies at $z=3.2-3.6$, we specifically select SFGs (and LBGs) from the VANDELS survey, ensuring to remove sources that have been flagged as AGN within VANDELS DR4. 
For the higher-redshift photometrically selected PRIMER$+$JADES sample at $z=6.9-7.6$, the potential contribution of AGN is expected to be minimal as a result of the rapid falloff in number density of AGN at $z\gtrsim5$ \citep[e.g.,][]{aird+15,Parsa18,mcgreer+18,kulkarni+19,faisst+21}. 

Although recent evidence points to a population of low-mass AGN at $z\approx5-7$ \citep{labbe+23}, these galaxies tend to be heavily dust-obscured (`little red dots') with a low ionizing output \citep{matthee+23,kocevski+23}, and are not expected to contribute significantly to the population of high-redshift {\oiiihb} emitters. However, to be conservative, all objects qualifying as little red dots were excluded when the PRIMER$+$JADES sample was selected (see Section \ref{sed:data_and_sample}). As a result, the underlying population of AGN, and its evolution over $z\simeq3-7$ are unlikely to significantly influence the {\EWoiiihb} evolution presented in this work.

\section{The ionizing photon production efficiency}\label{sec:xiion}

\begin{figure*}
    \centering
    \includegraphics[width=2.\columnwidth]{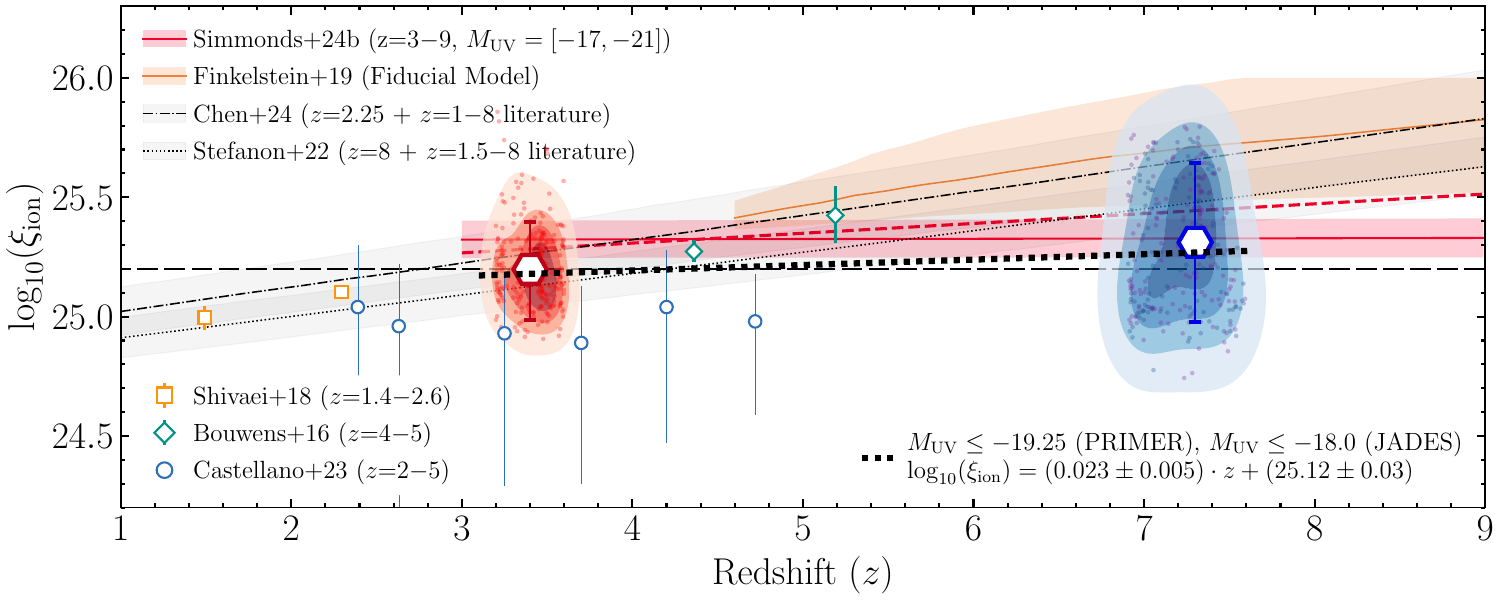}
    \includegraphics[width=2.\columnwidth]{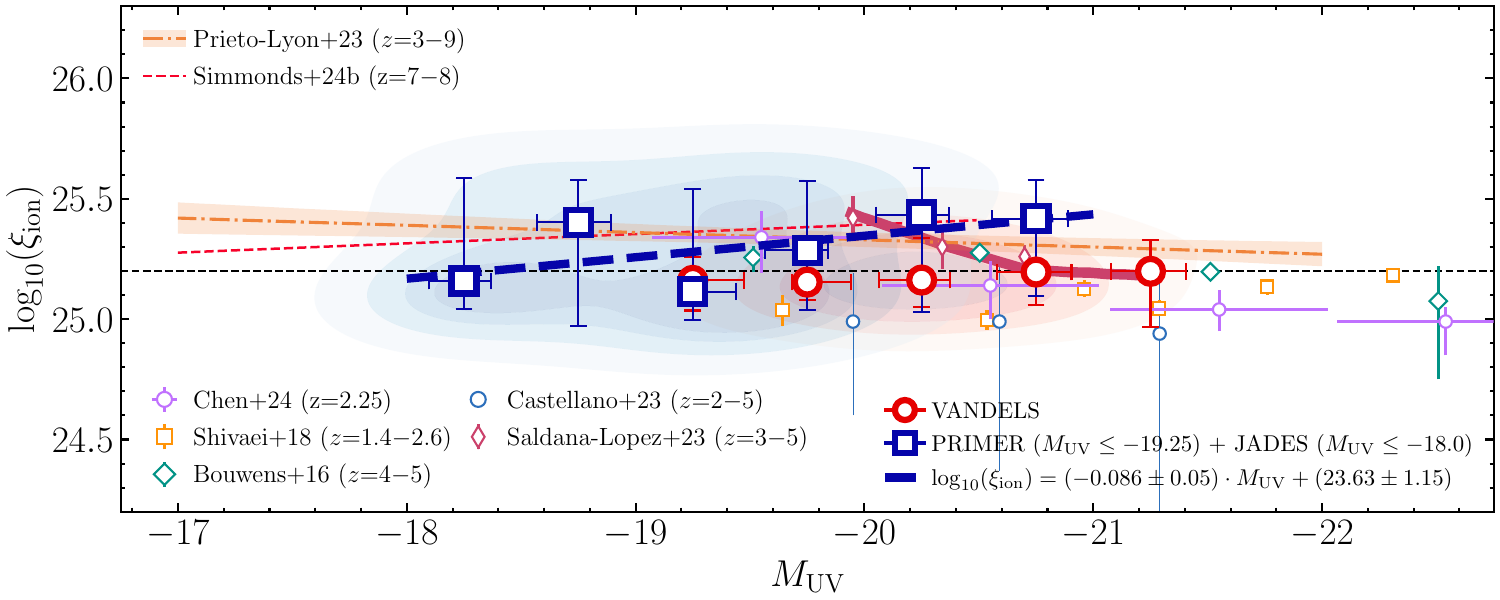}
    \caption{\textbf{(Top panel)} The ionizing photon production efficiency, {\xiion}, as a function of redshift. {\xiion} is inferred from our measured {\oiiihb} equivalent widths, with our $z=3.2-3.6$ VANDELS and photometrically complete ($>90\%$) $z=6.9-7.6$ PRIMER$+$JADES samples plotted as red and blue dots (with contours to aid visual clarity), respectively. The typical {\xiion} and scatter, as implied from the {\EWoiiihb} distribution median and width, are plotted as coloured hexagons. Fitting to the two samples in this work, we find a mild {\xiion$-z$} evolution over the redshift range $z\sim3-8$, given by $\mathrm{log}_{10}(\xi_\mathrm{ion} / \mathrm{Hz\, erg^{-1}})=(0.023\pm0.005)\cdot z + (25.12\pm0.03)$. The {\xiion} redshift evolution measurements by \citet{stefanon+22} (dotted black line) and \citet{chen+24} (dash-dotted black line) are plotted for comparison, in addition to the predicted {\xiion$-z$} relation from the reionization model presented by \citet{finkelstein+19}. The $\xi_\mathrm{ion}(z,M_{\mathrm{UV}})$ fit to a mass-complete sample using JADES by \citet{simmonds+24b} is plotted as a solid red line, with the surrounding red region indicating the scatter from galaxies having varying $M_\mathrm{UV}$ in the range $M_\mathrm{UV}\sim-19.0\pm2.0$. We also show the redshift evolution of the \citet{simmonds+24b} SFG-only sample as a red dashed line, and {\xiion} measurements from galaxy sample stacks across $z\sim1-5$ from \citet{bouwens+16,shivaei+18, castellano+23}. The horizontal dashed line equates to $\xi_\mathrm{ion}=10^{25.2}\,\mathrm{erg^{-1}\,Hz}$, the commonly used fiducial value in models for reionization to be completed by $z\sim5.5-6$ \citep[e.g.,][]{robertson+13,robertson+15}. Overall, this analysis, in conjunction with the available literature, supports a systematic $\approx2\times$ rise in the typical {\xiion} of SFGs from $z\sim2$ to $z\sim7.5$. \textbf{(Bottom panel)} {\xiion} as function of the absolute rest-frame UV magnitude, $M_\mathrm{UV}$ for the VANDELS and JWST-selected samples (represented as red and blue contours). The median and $\pm1\sigma$ {\xiion} in $\Delta(M_\mathrm{UV})=0.5\,$dex bins are shown for the JWST (blue squares) and VANDELS (red circles) samples. For our high-redshift sample, we find a mild $M_{\mathrm{UV}}-\xi_{\mathrm{ion}}$ anti-correlation, with a best-fitting relation given by $\mathrm{log}_{10}(\xi_\mathrm{ion} / \mathrm{Hz\, erg^{-1}})=(-0.086\pm0.05)\cdot M_\mathrm{UV} + (23.63\pm1.15)$. We find no evolution in {\xiion} with $M_\mathrm{UV}$ in our intermediate redshift VANDELS sample across $-19\lesssim M_\mathrm{UV}\lesssim-21.5$. In addition to the same stacking-based literature results plotted in the top panel, we also include results from \citet{saldana-lopez+23} and \citet{chen+24}, as well as the {\xiion$-M_\mathrm{UV}$} fit by \citet{prieto-lyon+23}.}
    \label{fig:xiion_fig_z_muv}
\end{figure*}

Motivated by the potential contribution of strong {\oiiihb} emitters to the ionizing photon budget, we investigate the ionizing photon production efficiency (\xiion) across our samples. To infer {\xiion}\footnote{Throughout this section, {\xiion} is the ionizing photon production efficiency assuming an escape fraction of $f_\mathrm{esc}=0$.}, we adopt the {\xiion$-W_\mathrm{\lambda}$([\mbox{O\,\sc{iii}}]$\lambda5007)$} relation presented in \citet{tang+19}: {$\mathrm{log}_{10}(\xi_\mathrm{ion} / \mathrm{Hz\, erg^{-1}})=0.76\times \mathrm{log}_{10}(W_\mathrm{\lambda}($[\mbox{O\,\sc{iii}}]$\lambda5007))+23.27$}, and a {\EWoiiihb$\rightarrow W_\mathrm{\lambda}($[\mbox{O\,\sc{iii}}]$\lambda5007))$} conversion factor of $\simeq0.67$ (see Section \ref{subsec:literature_z2}).

\subsection{Redshift evolution of {\xiion}}

The inferred {\xiion} for the VANDELS and PRIMER$+$JADES samples are shown in Fig. \ref{fig:xiion_fig_z_muv}, as a function of redshift, alongside literature measurements \citep[][]{bouwens+16,shivaei+18,finkelstein+19,stefanon+22,castellano+23,chen+24,simmonds+24b}. At $z=3.2-3.6$, our VANDELS sample galaxies have a median ionizing photon production efficiency of {$\mathrm{log}_{10}(\xi_\mathrm{ion} / \mathrm{Hz\, erg^{-1}})\simeq25.20$}, with a $1\sigma$ scatter of $\simeq0.2\,$dex. This is in excellent agreement with the predictions of \citet{stefanon+22} and \citet{chen+24}, who both estimate the {\xiion$-z$} evolution across a wide redshift range ($z\approx1-8$). Similarly, our measurements lie robustly on the redshift evolution expected by extrapolating between studies with samples at $z\sim2$ \citep{shivaei+18} and $z\sim5$ \citep{bouwens+16}. On the other hand, our inferred {\xiion} values are slightly above ($\Delta\xi_\mathrm{ion}\sim0.1-0.2\,$dex) those measured by \citet{castellano+23}. This is expected however, given that their redshift evolution is presented for a sample of VANDELS galaxies that is mass complete above $\mathrm{log}_{10}(M_*/\mathrm{M_\odot})\sim9.5$.

In our (photometrically complete) PRIMER$+$JADES sample ($z=6.9-7.6$), we find the median {\xiion} to be {$\mathrm{log}_{10}(\xi_\mathrm{ion} / \mathrm{Hz\, erg^{-1}})\simeq25.32$}, with the sample scatter increasing to $\simeq0.3\,$dex (as expected given the broader inferred {\EWoiiihb} distributions). 

In agreement with recent results based on JADES galaxies from \citet{simmonds+24} \citep[see also][]{simmonds+24b}, we find that the typical {\xiion} of our high-redshift sample is lower than previous results obtained prior to the availability of JWST imaging \citep[e.g.,][]{stefanon+22,chen+24}, as well as model predictions for high-redshift galaxy populations. Specifically, both the modelling of \citet{finkelstein+19} and the literature-compilation-based relation found by \citet{chen+24} suggest {$\Delta(\xi_\mathrm{ion})\sim0.2-0.4\,$dex} higher on average (albeit with large scatter), with the latter relation increasing as $\mathrm{d}\xi_{\mathrm{ion}}/\mathrm{d}z\simeq0.1$. In contrast, between our VANDELS $z=3.2-3.6$ and PRIMER$+$JADES $z=6.9-7.6$ samples, we find a relation given by: $\mathrm{log}_{10}(\xi_\mathrm{ion} / \mathrm{Hz\, erg^{-1}})=(0.023\pm0.005)\cdot z + (25.12\pm0.03)$, which is much shallower than previously expected \citep[a conclusion supported by][]{simmonds+24}.

This trend can be partially explained by sample selection effects impacting results prior to JWST, in which the galaxy samples observed were generally the brighter subset of the full population \citep{simmonds+24}. Another contributor to the shallower redshift evolution (and higher scatter) into the epoch of reionization is the increased prevalence of bursty star-formation, leading to a greater fraction of galaxies spending time in a star-formation lull with lower {\xiion} \citep[e.g.,][]{looser+23,dome+24,endsley+24,faisst+24,simmonds+24}.

We note that if we take both redshift and $M_\mathrm{UV}$ into account simultaneously and perform a multivariate fit, we recover the relation: $\mathrm{log}_{10}(\xi_\mathrm{ion} / \mathrm{Hz\, erg^{-1}})=(0.034\pm0.007)\cdot z - (0.037\pm0.014)\cdot M_\mathrm{UV} + (24.32\pm0.31)$. This relation is in broad agreement with other recent literature results, taking into account differences in sample selection and methodology \citep[e.g., see][]{simmonds+24,simmonds+24b,pahl+24}. Although steeper than when fitting for redshift evolution alone, this relation is still significantly flatter than many previous results in the literature.

\citet{pahl+24} have recently also published a study of the ionizing photon production efficiency of galaxies, in this case based on NIRSpec spectroscopy from the JADES and CEERS JWST surveys. While benefitting from spectroscopy for all sources (albeit for faint sources, {$W_\mathrm{\lambda}$} can often be more reliably determined from photometry) this study is limited by data of mixed quality and by small number statistics, especially when the sample is binned in redshift (their sample comprises 160 objects spanning the redshift range $1\leq z \leq6.7$). As such, the analysis presented in this work offers a unique vantage point from which to establish the redshift evolution of {\xiion} to high redshifts, whilst also uncovering the correlations between {\xiion} and physical properties (e.g., $M_\mathrm{UV}$, stellar mass) during a key phase of the epoch of reionization.

\begin{figure*}
    \centering
    \includegraphics[width=2.\columnwidth]{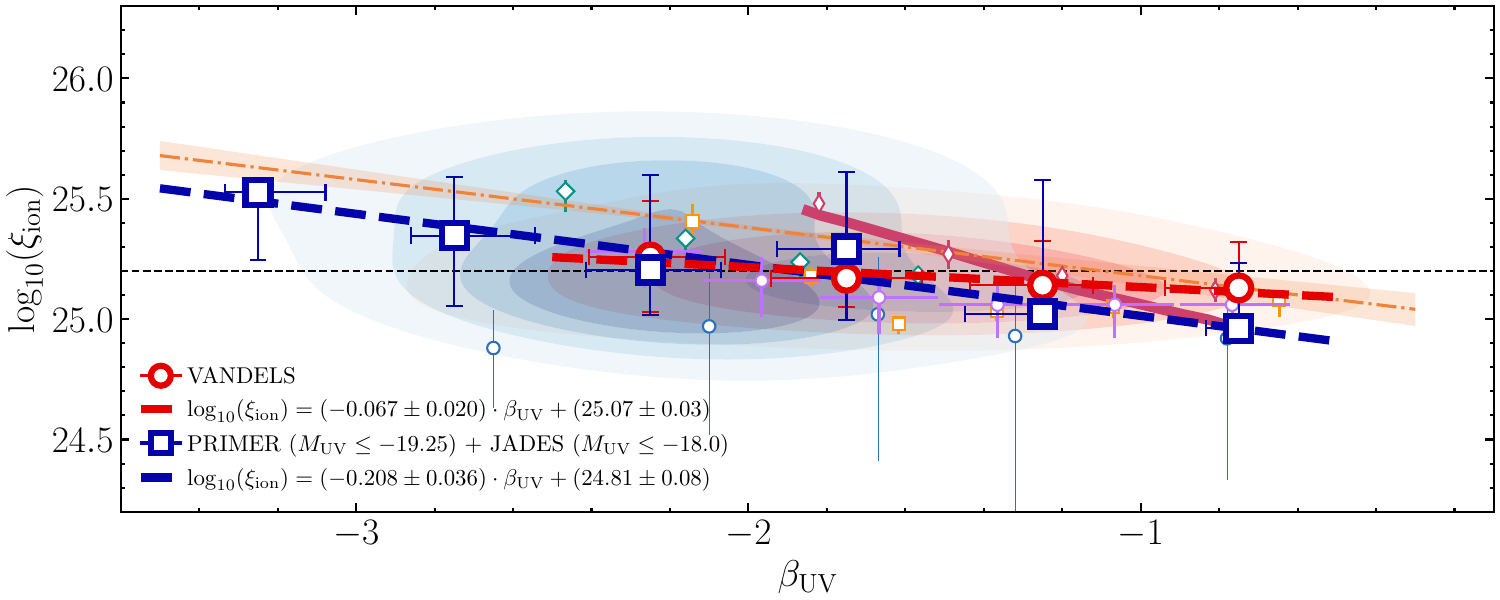}
    \includegraphics[width=2.\columnwidth]{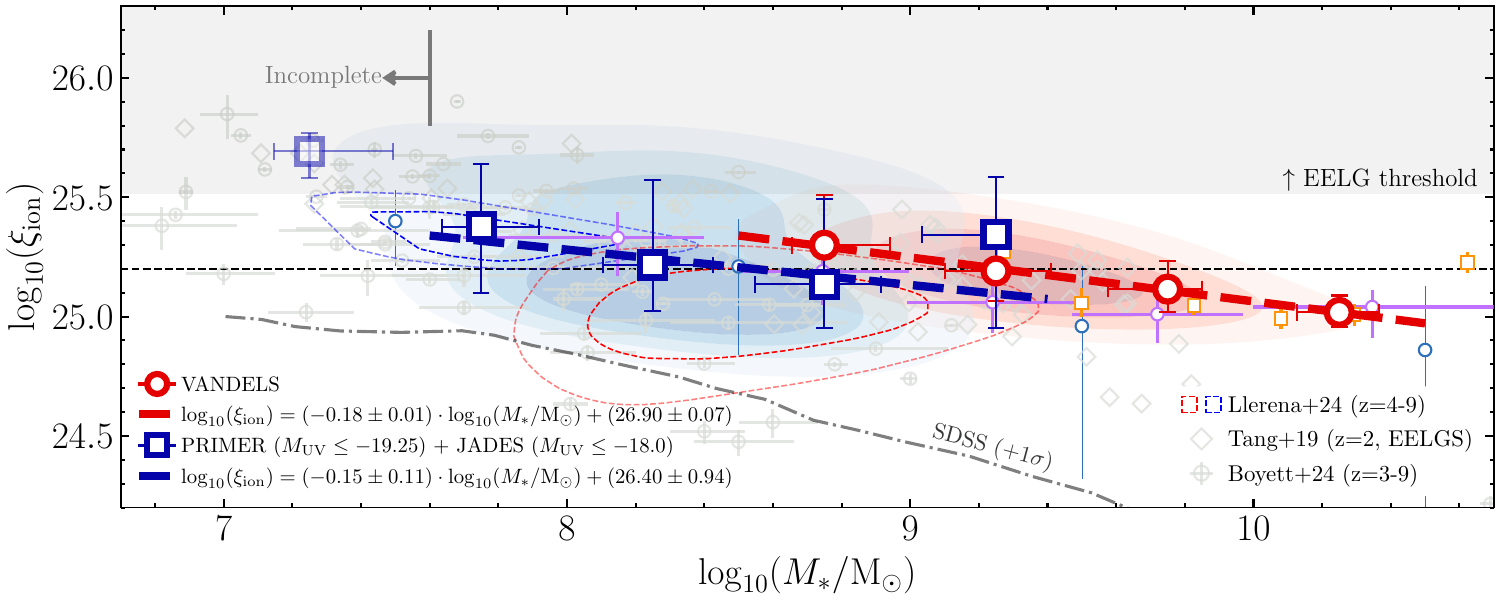}
    \caption{Same as Fig. \ref{fig:xiion_fig_z_muv}, but for the {\xiion} dependence on the UV continuum slope, $\beta_\mathrm{UV}$ (\textbf{upper}), and the stellar mass, $\mathrm{log}_{10}(M_*/M_\odot)$ (\textbf{lower}), with the large blue and red outlined markers representing the binned median {\xiion} and $\pm1\sigma$ values for the photometrically complete JWST ($\langle z\rangle\sim 7.3$) and VANDELS ($\langle z\rangle\sim 3.4$) samples, respectively. In addition to the stacking-based literature results in Fig. \ref{fig:xiion_fig_z_muv}, we include the {\xiion$-\beta_{\mathrm{UV}}$} relation derived by \citet{prieto-lyon+23} (orange dashed line) using the UNCOVER and GLASS JWST surveys as well as contours (dashed red/blue lines) for the CEERS-selected samples from \citet{llerena+24}. Both  $\beta_{\mathrm{UV}}$ and $\mathrm{log}_{10}(M_*/\mathrm{M_\odot})$ anti-correlate with {\xiion} described by best-fitting relations shown in the bottom-left of each panel. Based on our photometric completeness $M_{\mathrm{UV}}$ limits, we are approximately mass-complete at $\mathrm{log_{10}(M_*/M_\odot)}\gtrsim7.6$. For a comparison with the properties of local galaxies, we plot the upper $1\sigma$ ($84^{\mathrm{th}}$ percentile) {\xiion$-\mathrm{log}_{10}(M_*/M_\odot)$} relation from the SDSS \citep[see][converted using the \citealt{tang+19} relation]{matthee+23}. We note that from the \citet{tang+19} relation used to estimate {\xiion} from our {\EWoiiihb} measurements, EELGs ({\EWoiiihb$\ >1000\,$\AA}) have {\xiion$\,\gtrsim10^{25.5}\,\mathrm{erg^{-1}\,Hz}$}, marked as the grey shaded region.} 
    \label{fig:xiion_fig_beta_mass}
\end{figure*}

In agreement with the trends seen in Fig. 8, \citet{pahl+24} unveil a similarly mild but significant positive evolution in {\xiion} with redshift, and between {\xiion} and UV luminosity (at marginal significance, although primarily driven by their data at  $z\sim2$). However, due to the statistical limitations (compounded by the broad spread in redshift and resulting correlation of $M_\mathrm{UV}$ with redshift), \citet{pahl+24} do not reveal the relation found here between {\xiion} and UV slope (Fig. \ref{fig:xiion_fig_z_muv}\,a), and the clear and crucial negative relation between {\xiion} and stellar mass revealed in the present study (Fig. \ref{fig:xiion_fig_beta_mass}\,b).

\subsection{Dependence on $M_\mathrm{UV}$}

The ionizing photon production efficiency as a function of absolute UV magnitude is shown in Fig. \ref{fig:xiion_fig_z_muv} (bottom) panel, demonstrating a mild anti-correlation in our JWST-selected sample, given as $\mathrm{log}_{10}(\xi_\mathrm{ion} / \mathrm{Hz\, erg^{-1}})= (-0.086\pm0.05)\cdot M_\mathrm{UV} + (23638\pm1.15)$.
Although this relation is evident in the high-redshift sample, with brighter $M_\mathrm{UV}$ galaxies showing higher {\xiion} (increased {\EWoiiihb}; in agreement with \citet{pahl+24}), the {\xiion$-M_\mathrm{UV}$} trend is not clear when taking the full redshift and $M_\mathrm{UV}$ parameter space into account as the VANDELS sample shows no relation with $M_\mathrm{UV}$. This is consistent with the null correlations seen in the recent literature \citep[e.g.,][]{bouwens+16,shivaei+18,castellano+23}. Given that some works reveal mild (and on occasion, opposing) trends \citep[e.g., see][]{prieto-lyon+23,saldana-lopez+23,simmonds+24} between {\xiion} and $M_\mathrm{UV}$, it is clear that other factors that change with redshift are more dominant \citep[e.g., evolving SFH, metallicities, etc.;][]{cullen+19,endsley+24}. This conclusion is supported by the significant evolution observed in the typical {\EWoiiihb} at fixed {$M_\mathrm{UV}\simeq-19.8$} between our samples shown in Fig. \ref{fig:ew0_distn_physicalparametersplit}. It is clear that to fully establish any trends between {\xiion} and {$M_\mathrm{UV}$}, and indeed if these evolve with redshift,  larger samples at both intermediate and high redshifts, with increased dynamic ranges in UV luminosity, will be required.

\subsection{Dependence on $\mathbf{\beta_\mathrm{UV}}$}
Fig. \ref{fig:xiion_fig_beta_mass} (top panel) shows the observed {\xiion$-\beta_{\mathrm{UV}}$} anti-correlation observed in our analysis. Fitting a linear relation to our JWST data, we find: $\mathrm{log}_{10}(\xi_\mathrm{ion} / \mathrm{Hz\, erg^{-1}})=(-0.208\pm0.036)\cdot \beta_\mathrm{UV} + (24.81\pm0.08)$. In the VANDELS sample, we find a similar but slightly weaker relation given by: $\mathrm{log}_{10}(\xi_\mathrm{ion} / \mathrm{Hz\, erg^{-1}})=(-0.067\pm0.020)\cdot \beta_\mathrm{UV} + (25.07\pm0.03)$. These relations are in broad agreement with existing literature \citep[e.g., see][]{prieto-lyon+23}, although slightly shallower than found in \citet{saldana-lopez+23}. In the former, the sample is constructed from $\mathrm{Ly\alpha}$-detected MUSE sources, and is likely a more extreme sub-population of galaxies, whilst in the latter, the steeper slope may be attributed to differences in the measured UV continuum slopes in photometric versus spectroscopic data \citep[e.g., see Section 6 in;][see also \citealt{calabro+22}]{saldana-lopez+23}.

\vspace{-0.2 cm}
\subsection{Dependence on $\mathrm{log}_{10}(M_*/\mathrm{M_\odot})$}

Lastly, in the bottom panel of Fig. \ref{fig:xiion_fig_beta_mass}, we plot the trends of {\xiion} against inferred stellar mass. Across both the samples studied in this work, the trends with stellar mass are the least ambiguous of those investigated, with a relationship across our full dataset of: $\mathrm{log}_{10}(\xi_\mathrm{ion} / \mathrm{Hz\, erg^{-1}})=(-0.15\pm0.09)\cdot \mathrm{log}_{10}(M_*/\mathrm{M_\odot}) + (26.40\pm0.94)$ in our PRIMER$+$JADES sample and $\mathrm{log}_{10}(\xi_\mathrm{ion} / \mathrm{Hz\, erg^{-1}})=(-0.18\pm0.01)\cdot \mathrm{log}_{10}(M_*/\mathrm{M_\odot}) + (26.90\pm0.07)$ in our VANDELS sample. The fitted relations here are consistent within $\pm1\sigma$ and agree remarkably well with literature results over a wide range of redshifts and a wide range of sample selections, including both photometric and spectroscopic studies \citep[][]{shivaei+18,tang+19,castellano+23,boyett+24,chen+24,llerena+24}. Generally speaking, there appears to be a relatively fundamental {\xiion$-\mathrm{log}_{10}(M_*/\mathrm{M_\odot})$} relation from cosmic noon to the reionization epoch, sitting significantly above the relation seen in the local Universe \citep[e.g., using SDSS,][]{matthee+23}. We note that given the photometric selection described in Section \ref{subsubsec:final_sample}, our sample will be impacted by stellar mass incompleteness below $\mathrm{log_{10}(M_*/M_\odot})\sim7$ and these galaxies are therefore excluded from the best-fitting relation (see also Section \ref{subsec:distribution_with_properties}).

\subsection{Implications for the epoch of reionization}
A significant fraction of the lower-mass SFG population ($\mathrm{log}_{10}(M_*/\mathrm{M_\odot})\lesssim8.0$) display extreme optical emission-line equivalent widths (\EWoiiihb\ $\gtrsim1000\,$\AA). Such galaxies likely have lower metallicities, relatively bluer UV slopes and are undergoing an intense surge in star-formation, all of which becomes more common in the burstier population at these redshifts. In turn, it is expected that these galaxies represent a substantial subset of those that dominate the ionizing photon budget and in turn are likely the primary drivers of reionization.

Importantly for the discussion of how reionization progressed, in addition to the presence of galaxies with relatively extreme {\xiion} (e.g., {$\gtrsim10^{25.6}\,\mathrm{erg^{-1}Hz}$}), we see a significant fraction, $\sim0.44$ ($\sim0.13$), of galaxies with {\xiion} below the typical canonical value in reionization simulations \citep[e.g.,][]{robertson+15}, $\mathrm{log}_{10}(\xi_\mathrm{ion} / \mathrm{Hz\, erg^{-1}})\leq25.2\, (25.0)$. 

Moreover, assuming the $\beta_{\mathrm{UV}}-f_\mathrm{esc}$ relation from \citet{chisholm+22} applies in EOR galaxy populations, our sample displays a more moderate typical LyC escape fraction of $f_\mathrm{esc}\simeq5\,$per cent. Indeed, only $\sim11\,(22)\,$per cent of the galaxies in our PRIMER$+$JADES samples have $f_\mathrm{esc}\geq10\,$per cent and $\mathrm{log}_{10}(\xi_\mathrm{ion} / \mathrm{Hz\, erg^{-1}})\geq 25.5\, (25.2)$. Therefore, it is clear not all galaxies at high redshift are producing, and leaking, extreme amounts of ionizing photons, and are likely more moderate in their output \citep[in part due to the increased time variability in {\fesclyc} and {\xiion} as a result of burstier star-formation modes;][]{maji+22,endsley+24}. 

In conclusion, there appears to be sufficient photons to drive reionization, but not so many that there is a `budget crisis' \citep[e.g.,][]{munoz+24} in which reionization would finish too early as a result of an over-production of ionizing photons.

\section{Conclusions}\label{sec:conclusions}

In this work we have used JWST/NIRCam imaging from the PRIMER and JADES surveys to identify a sample of $N=289$ galaxies during a key phase within the Epoch of Reionization ($6.9<z<7.6$). For this sample, we have made robust estimates of their {\oiiihb} emission-line equivalent widths as a probe of their ionizing properties, and, importantly, how these measurements vary with key physical properties such as their absolute UV magnitudes ($M_\mathrm{UV}$), UV continuum slopes ($\beta_\mathrm{UV}$) and stellar masses. We supplement this high-redshift dataset with a sample of galaxies at $3.2<z<3.6$ selected from the VANDELS spectroscopic survey, providing an important benchmark for the ionizing properties of galaxies at intermediate redshifts between cosmic noon and the EOR. The main results of our analysis can be summarised as follows:

\begin{enumerate}
    \item We find a clear redshift evolution in the {\EWoiiihb} distribution between our two samples (see Fig. \ref{fig:ew0_distn_main}), with the median {\EWoiiihb} increasing by a factor of $\simeq1.5\times$ from {\EWoiiihb}\,$=380\pm18\,$\AA\, in our VANDELS sample to {\EWoiiihb}\,$=540\pm25\,$\AA\, in our JWST-selected sample. This evolution can be approximately described by a power-law in time with $\alpha\sim-0.5$. This is consistent with the higher-redshift galaxy population having lower metallicities \citep{cullen+19} and younger stellar populations, which are more ionizing and produce stronger {\oiiihb} emission \citep{endsley+24}.

    \item A clear broadening ($\sim0.16\,$dex) in the {\EWoiiihb} distribution is present for our high-redshift sample relative to that seen in VANDELS, in addition to a clear departure from the log-normal functional form. This increased width, and more prominent tail of galaxies with lower {\EWoiiihb}, is consistent with the high-redshift galaxy population having `burstier' star-formation histories \citep{endsley+24,langeroodi+24}, with phases of intense star-formation being intermittent with subsequent lulls, a process regulated by the gas duty cycle \citep[e.g.,][]{looser+23,dome+24,witten+24}.

    \item By establishing how the {\EWoiiihb} distribution evolves between different subsamples, we find that $M_\mathrm{UV}$-faint galaxies at high redshift, and those with redder UV slopes, have systematically weaker {\oiiihb} emission. In contrast, we find that the lower-mass dwarf SFGs ($\mathrm{log}_{10}(M_*/\mathrm{M_\odot})\lesssim8.5$) have the highest {\EWoiiihb} as a result of the clear {\EWoiiihb$\mathrm{log}_{10}-(M_*/\mathrm{M_\odot})$} anti-correlation (see Fig. \ref{fig:ew0_distn_physicalparametersplit}). This mass dependence appears to be approximately redshift invariant beyond cosmic noon \citep[][]{matthee+22}.

    \item Taking our analysis in conjunction with lower-redshift emission-line galaxy studies \citep[e.g., $z\sim2$;][]{boyett+22}, our results constitute strengthened evidence for a systematic increase in the prevalence of EELGs (\EWoiiihb$\gtrsim1000\,$\AA) with redshift. Overall, the EELG fraction increases from $\sim4\,$per cent at $z\sim2$, to $\simeq8^{+1}_{-1} $per cent in our VANDELS sample and $\sim29{\pm2}\,$per cent in the PRIMER$+$JADES sample.

    \item Using empirically established relations between the ionizing photon production efficiency and {\oiii} \citep[e.g.,][]{tang+19}, we infer {\xiion} for our samples and measure a milder redshift evolution than previously estimated from pre-JWST studies \citep{finkelstein+19,stefanon+22,chen+24}: $\mathrm{log}_{10}(\xi_\mathrm{ion} / \mathrm{Hz\, erg^{-1}})~=~(0.023~\pm~0.005)~\cdot z + (25.12\pm0.03)$, in agreement with \citet{simmonds+24}. The milder observed redshift evolution, driven by a non-negligible fraction of galaxies with less extreme {\xiion} ($\lesssim 10^{25.2}\,\mathrm{erg^{-1}Hz}$) and more scatter, alleviates concerns about a `budget crisis', whereby the high-redshift galaxy population produces too many photons during reionization \citep{munoz+24}.

    \item As shown in Fig. \ref{fig:xiion_fig_z_muv} and Fig. \ref{fig:xiion_fig_beta_mass}, low-mass ($\mathrm{log}_{10}(M_*/\mathrm{M_\odot})\lesssim8$) and relatively UV-bright ($M_\mathrm{UV}\lesssim-19$) galaxies with bluer UV continua ($\beta\lesssim-2.2$) tend to have higher ionizing photon production efficiencies. On that account, this sub-population of galaxies at $z\gtrsim6$ are expected to be the dominant drivers of reionization.
    
\end{enumerate}


\section*{Acknowledgements}
R. Begley, R. J. McLure, J. S. Dunlop, D.J. McLeod, and C. Donnan acknowledge the support of the Science and Technology Facilities Council. F. Cullen and T. M. Stanton acknowledge the support from a UKRI Frontier Research Guarantee Grant [grant reference EP/X021025/1]. A. C. Carnall acknowledges support from a UKRI Frontier Research Guarantee Grant [grant reference EP/Y037065/1]. JSD acknowledges the support of the Royal Society via the award of a Royal Society Research Professorship. RSE acknowledges generous financial support from the Peter and Patricia Gruber Foundation. RKC is grateful for support from the Leverhulme Trust via the Leverhulme Early Career Fellowship.

This work is based in part on observations made with the
NASA/ESA/CSA James Webb Space Telescope . The data were obtained from the Mikulski Archive for Space Telescopes at the Space Telescope Science Institute, which is operated by the Association of Universities for Research in Astronomy, Inc., under NASA contract NAS 5-03127 for JWST. The authors acknowledge the associated teams for
developing their observing programs with a zero-exclusive-access
period. This work also utilizes data from the JADES DR2$+$DR3 data release
(DOI: 10.17909/8tdj-8n28; \citealt{eisenstein+23a}; \citealt{rieke+23}; \citealt{deugenio+24} ). Some of the data products presented herein were retrieved from the Dawn \emph{JWST} Archive (DJA). DJA is an initiative of the Cosmic Dawn Center, which is funded by the Danish National Research Foundation under grant No. 140.

This research made use of Astropy, a community-developed core Python package for Astronomy \citep{astropy13,astropy18},  NumPy \citep{numpy20} and SciPy \citep{scipy20}, Matplotlib \citep{matplotlib07}, IPython \citep{ipython07} and NASA’s Astrophysics Data System Bibliographic Services.

\section*{Data Availability}
The VANDELS survey is a European Southern Observatory Public Spectroscopic Survey. The full spectroscopic dataset, together with the complementary photometric information and derived quantities are available from \url{http://vandels.inaf.it}, as well as from the ESO archive \url{https://www.eso.org/qi/}.\\
For the purpose of open access, the author has applied a Creative Commons Attribution (CC BY) licence to any Author Accepted Manuscript version arising from this submission.



\bibliographystyle{mnras}
\bibliography{oiiihb} 



\appendix

\section{\textsc{bagpipes} SED modelling assumptions}\label{appendix:bagpipes_assumptions}

With the aim of obtaining robust {\EWoiiihb} measurements from our \textsc{bagpipes} SED fitting, we tested multiple configurations of star-formation history, dust attenuation law, and stellar population models.
For star-formation histories, we explored the following models prescriptions: constant, delayed$-\tau$, continuity, bursty continuity, two-component constant (with the boundary fixed at $10\,\mathrm{Myr}$), and a two-component constant model with the boundary as a free parameter between $3\,\mathrm{Myr}$ and $10\,\mathrm{Myr}$. For the dust attenuation, we have tested a Calzetti dust attenuation law, a steeper SMC-like extinction law, and the flexible \citet{salim+18} prescription (parameterised by $\delta$, which acts to flatten or steepen the dust attenuation law). Lastly, we perform \textsc{bagpipes} runs with both BC03 and BPASS SPS models. A comparison of how well a subset of the configuration permutations perform in recovering the spectroscopic {\EWoiiihb} measurements is shown for the high-redshift sample in Fig. A1, and for the NIRVANDELS sample in Fig. A2. As discussed in Sections \ref{subsubsec:spec_measurements_nirv} and \ref{subsubsec:spec_measurements_jwst}, the fiducial \textsc{bagpipes} configurations used for our analysis are those that best recover the true {\EWoiiihb} values.

\begin{figure*}
    \begin{tabular}{cc}
         \subfloat[Constant SFH, Calzetti dust attenuation law, BPASS SPS models]{\includegraphics[width=1.0\columnwidth]{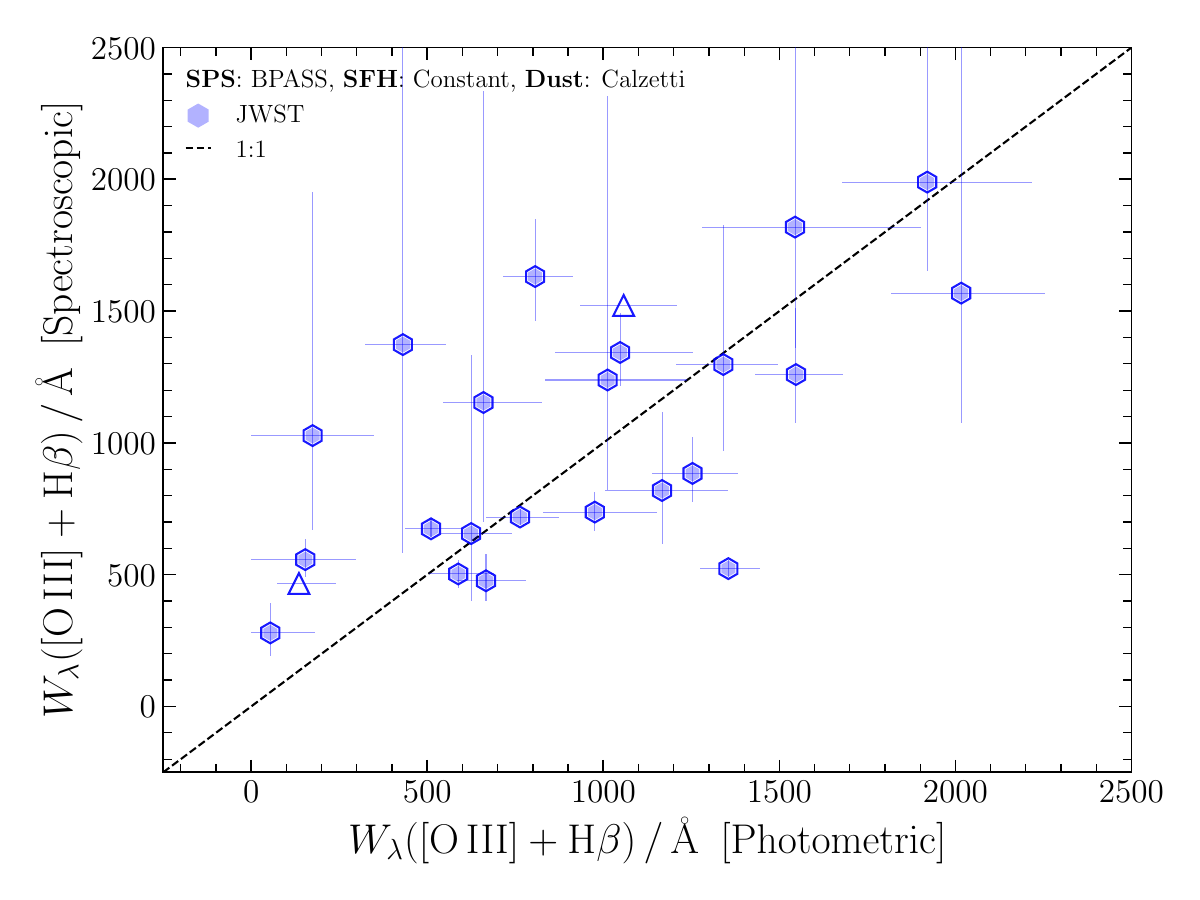}} & \subfloat[Continuity SFH, Calzetti dust attenuation law, BPASS SPS models]{\includegraphics[width=1.0\columnwidth]{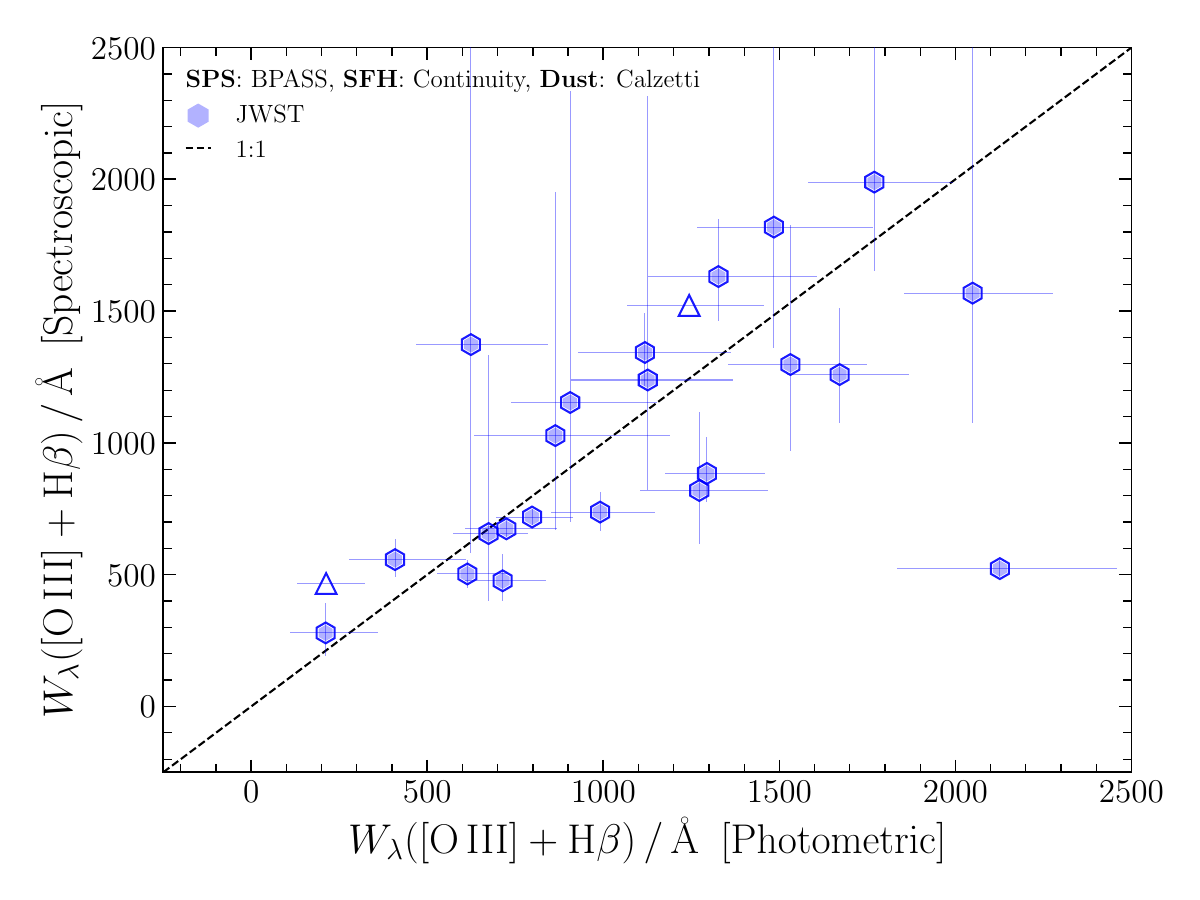}} \\ \subfloat[Bursty continuity SFH, Calzetti dust attenuation law, BPASS SPS models]{\includegraphics[width=1.0\columnwidth]{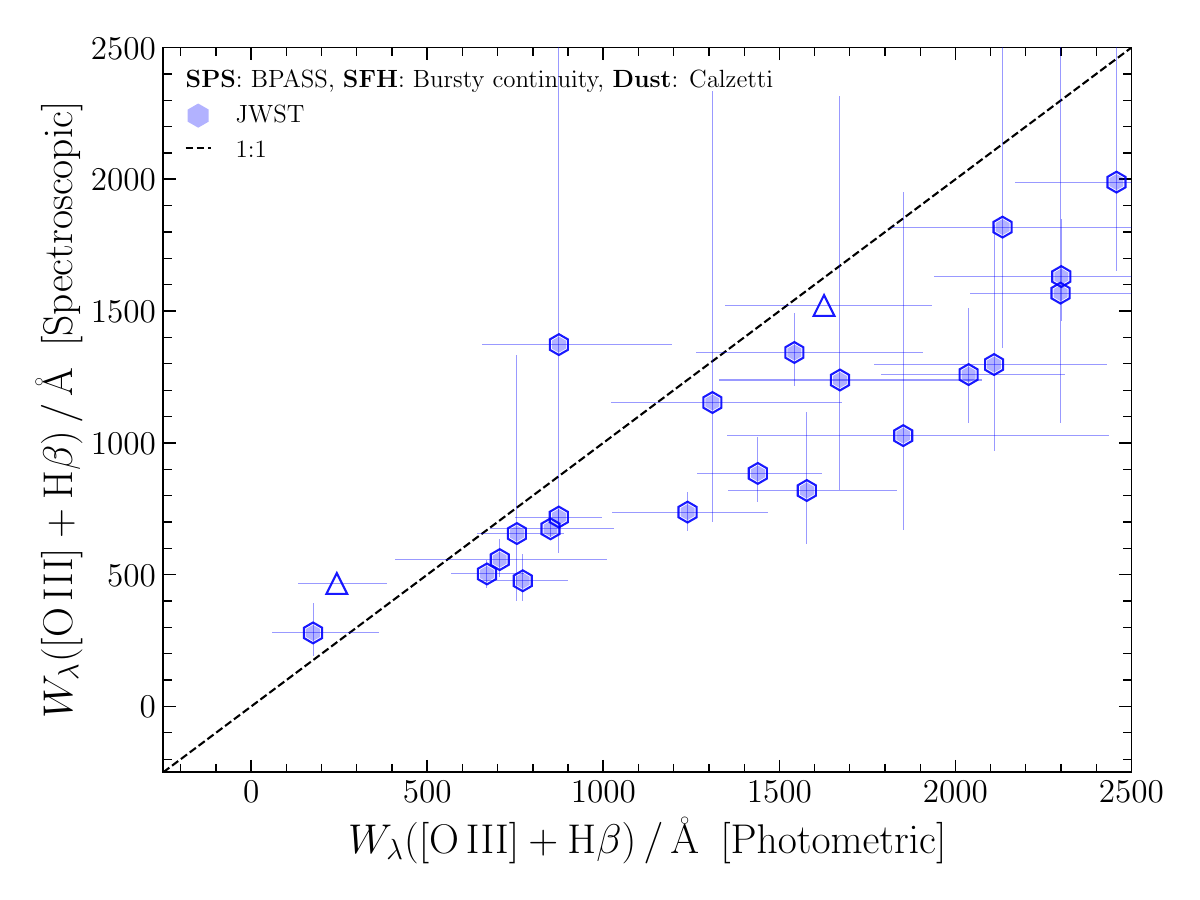}} 
         & \subfloat[Delayed$-\tau$ SFH, Calzetti dust attenuation law, BPASS SPS models]{\includegraphics[width=1.0\columnwidth]{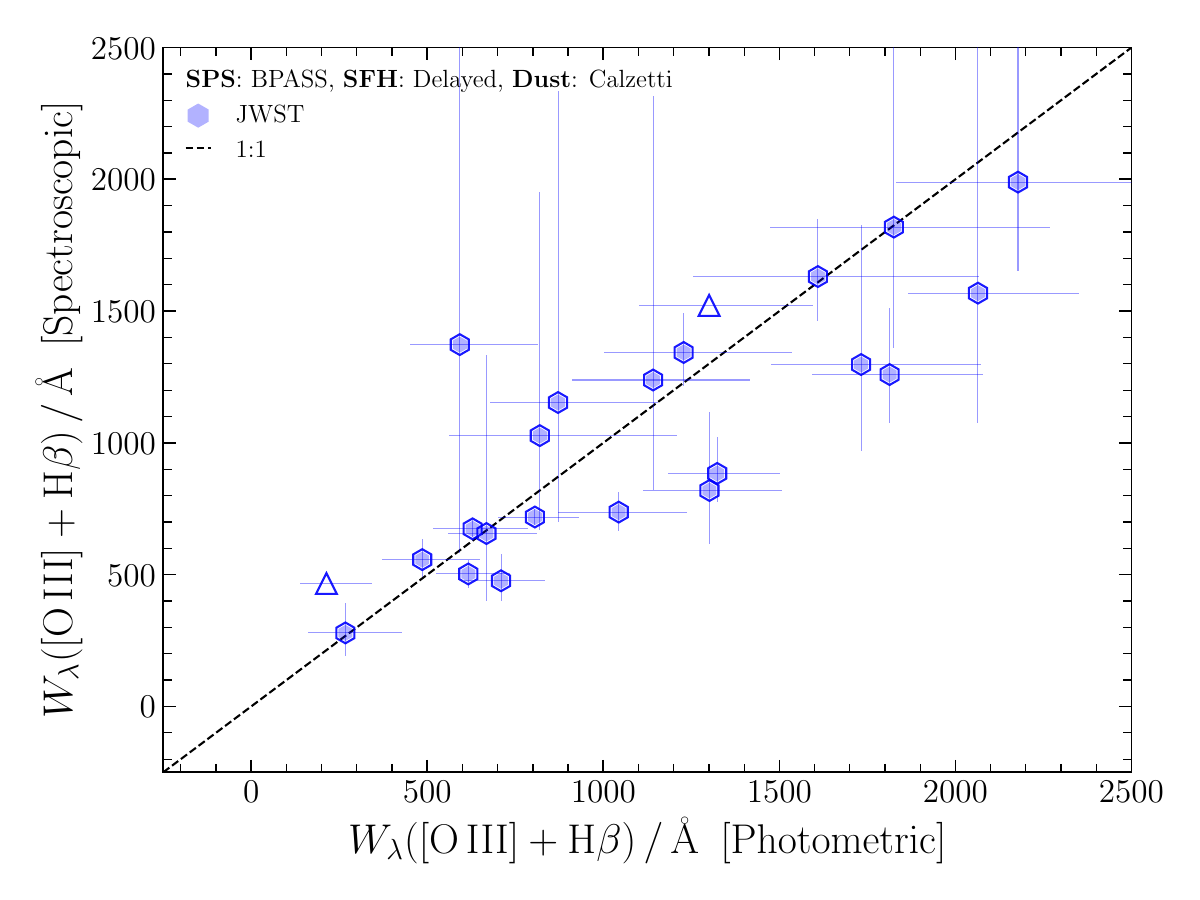}} \\ \subfloat[Delayed$-\tau$ SFH, SMC dust attenuation law, BPASS SPS models]{\includegraphics[width=1.0\columnwidth]{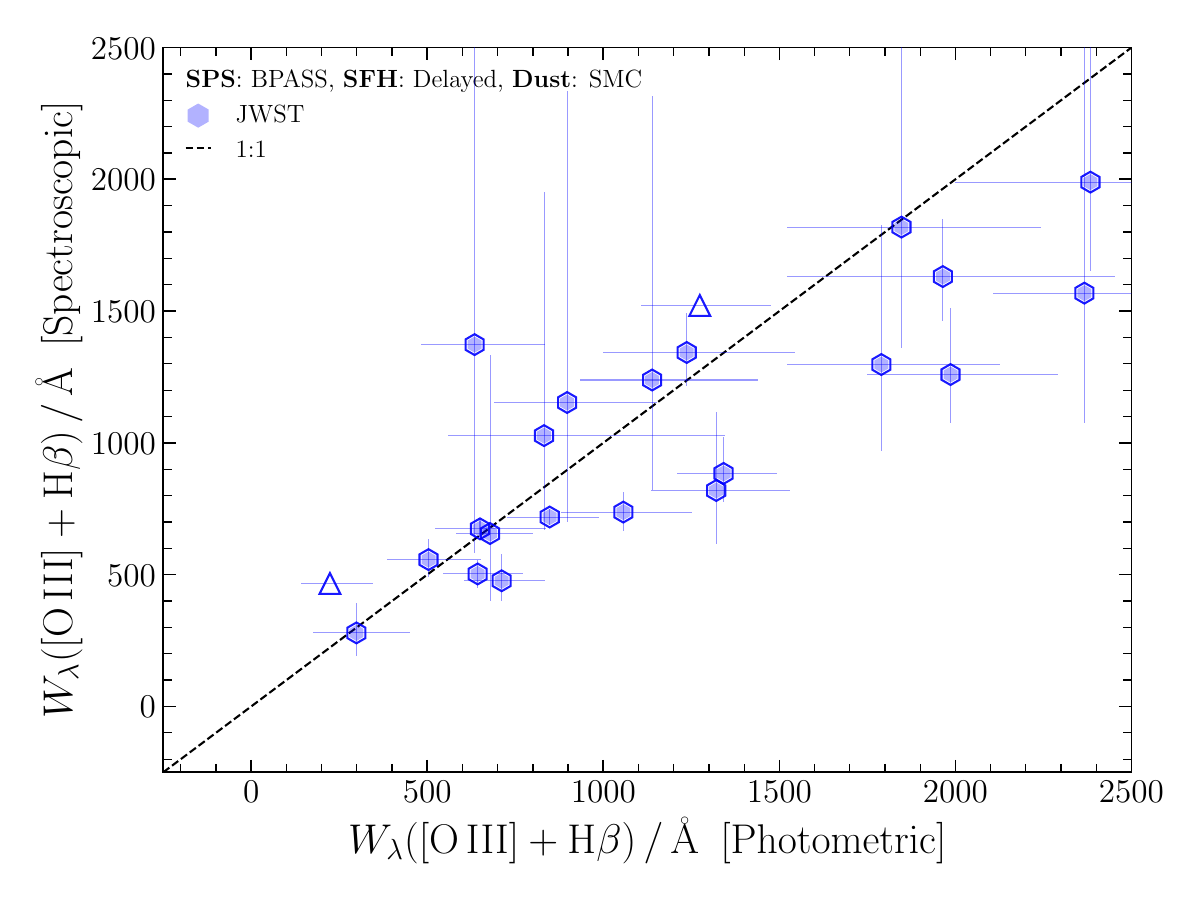}} & \subfloat[Continuity SFH, Calzetti dust attenuation law, BC03 SPS models]{\includegraphics[width=1.0\columnwidth]{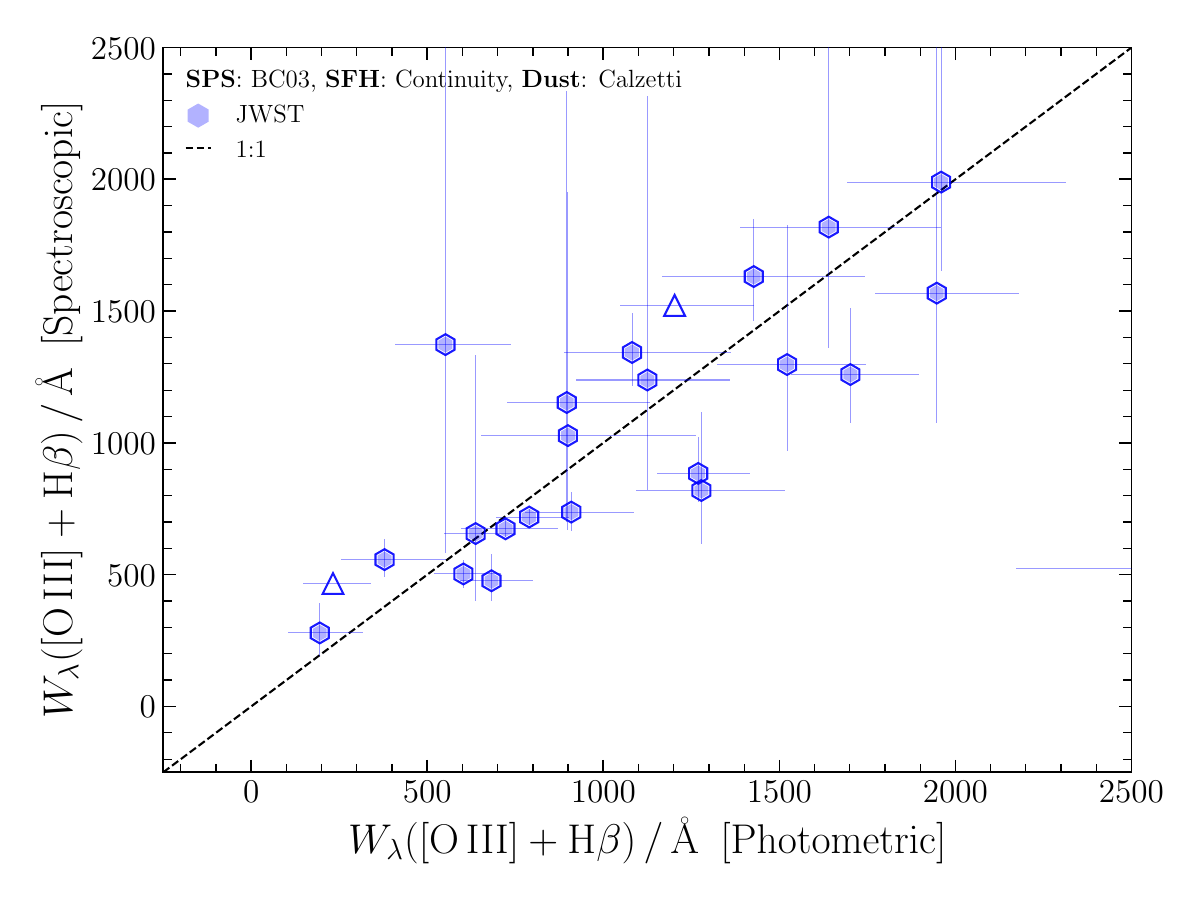}} 
    \end{tabular}
    \caption{Performance of different \textsc{bagpipes} configurations in recovering spectroscopic {\EWoiiihb} values from photometric SED fits for the high-redshift JWST spectroscopic sample.}
    \label{fig:appendix_A_jwst}
\end{figure*}

\begin{figure*}
    \begin{tabular}{cc}
         \subfloat[Constant SFH, Calzetti dust attenuation law, BPASS SPS models]{\includegraphics[width=1.0\columnwidth]{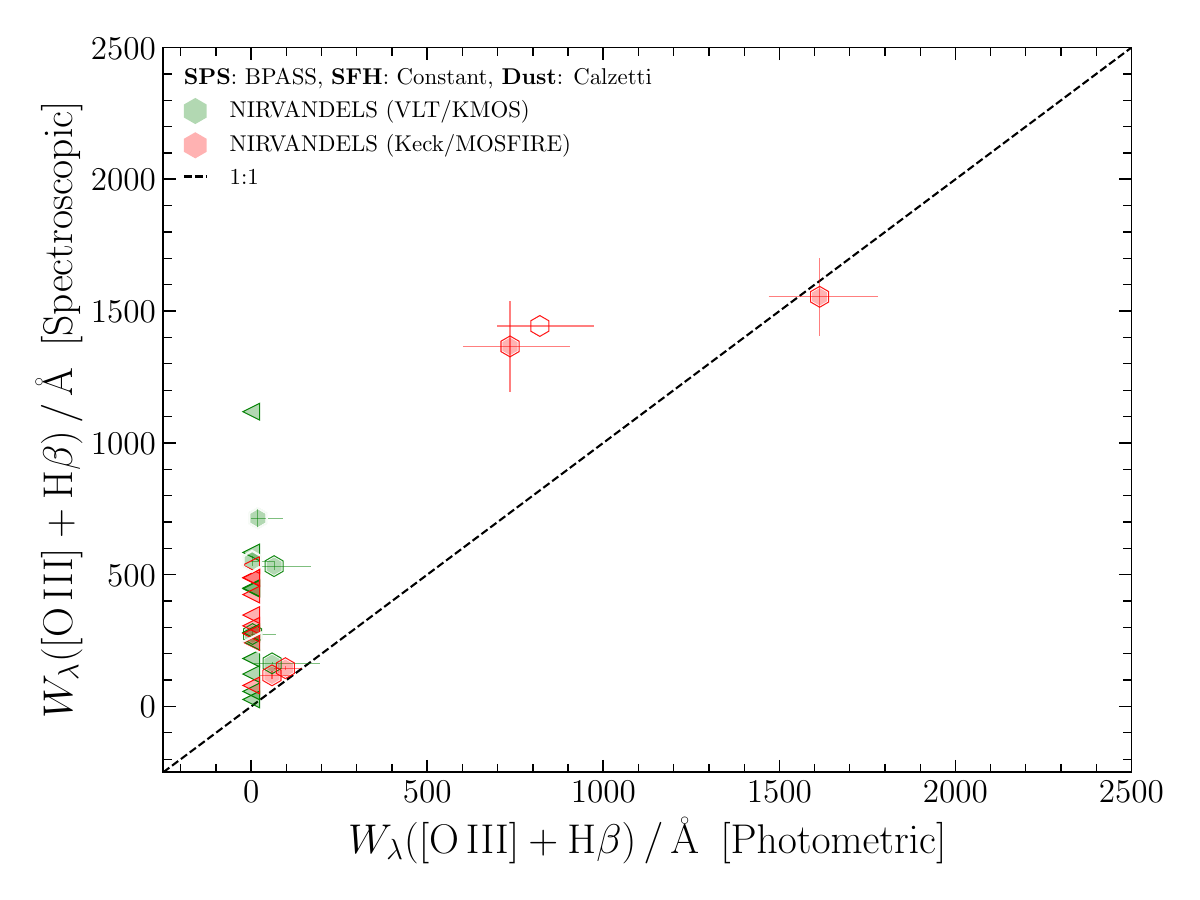}} & \subfloat[Continuity SFH, Calzetti dust attenuation law, BPASS SPS models]{\includegraphics[width=1.0\columnwidth]{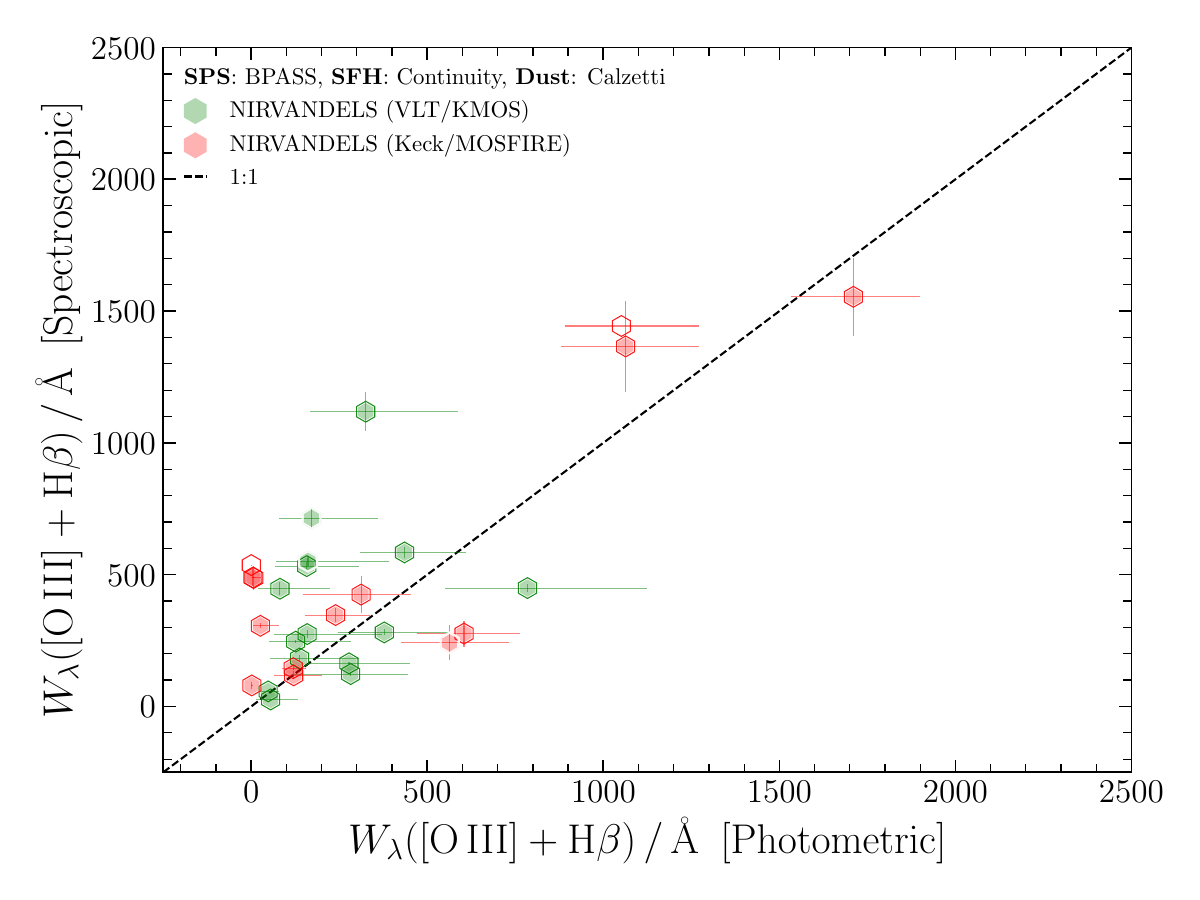}} \\ \subfloat[Bursty continuity SFH, Calzetti dust attenuation law, BPASS SPS models]{\includegraphics[width=1.0\columnwidth]{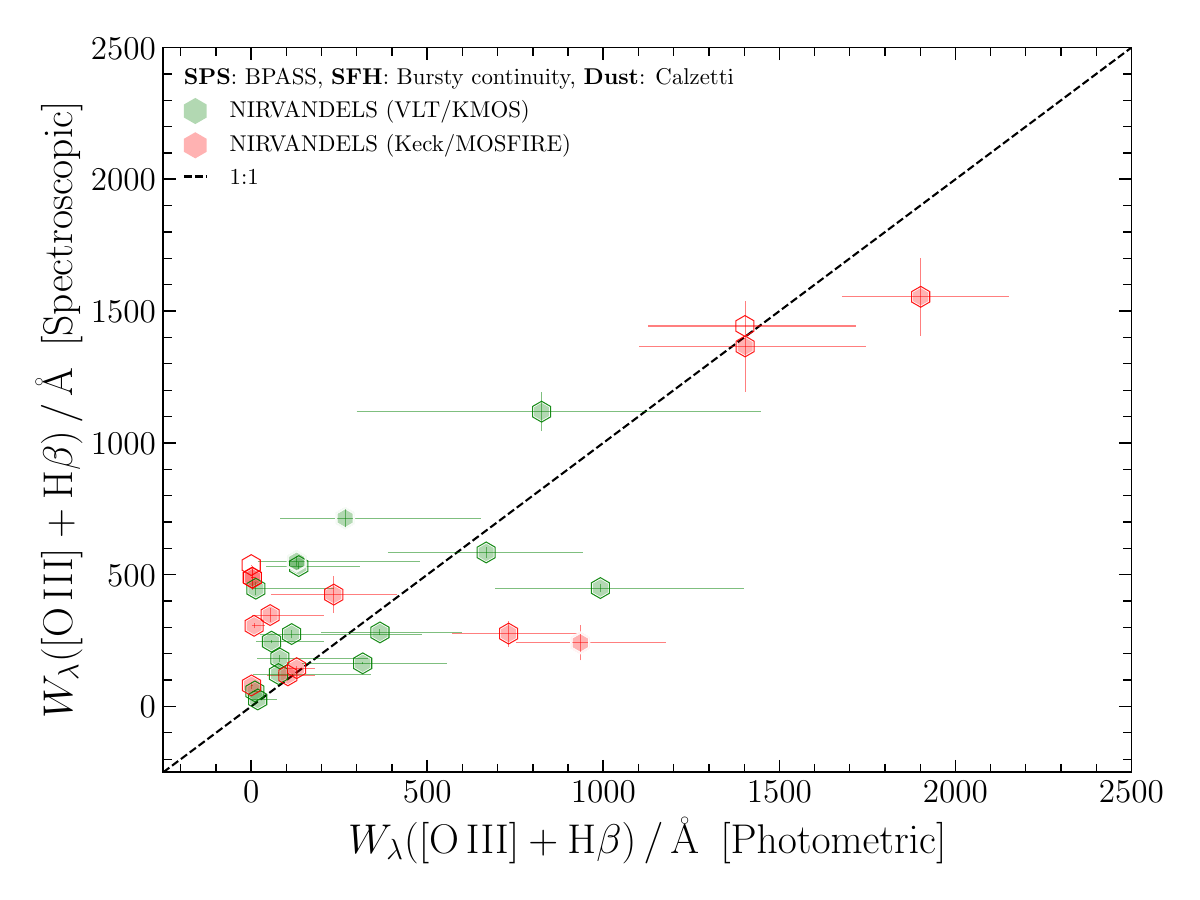}} & \subfloat[Delayed$-\tau$ SFH, Calzetti dust attenuation law, BPASS SPS models]{\includegraphics[width=1.0\columnwidth]{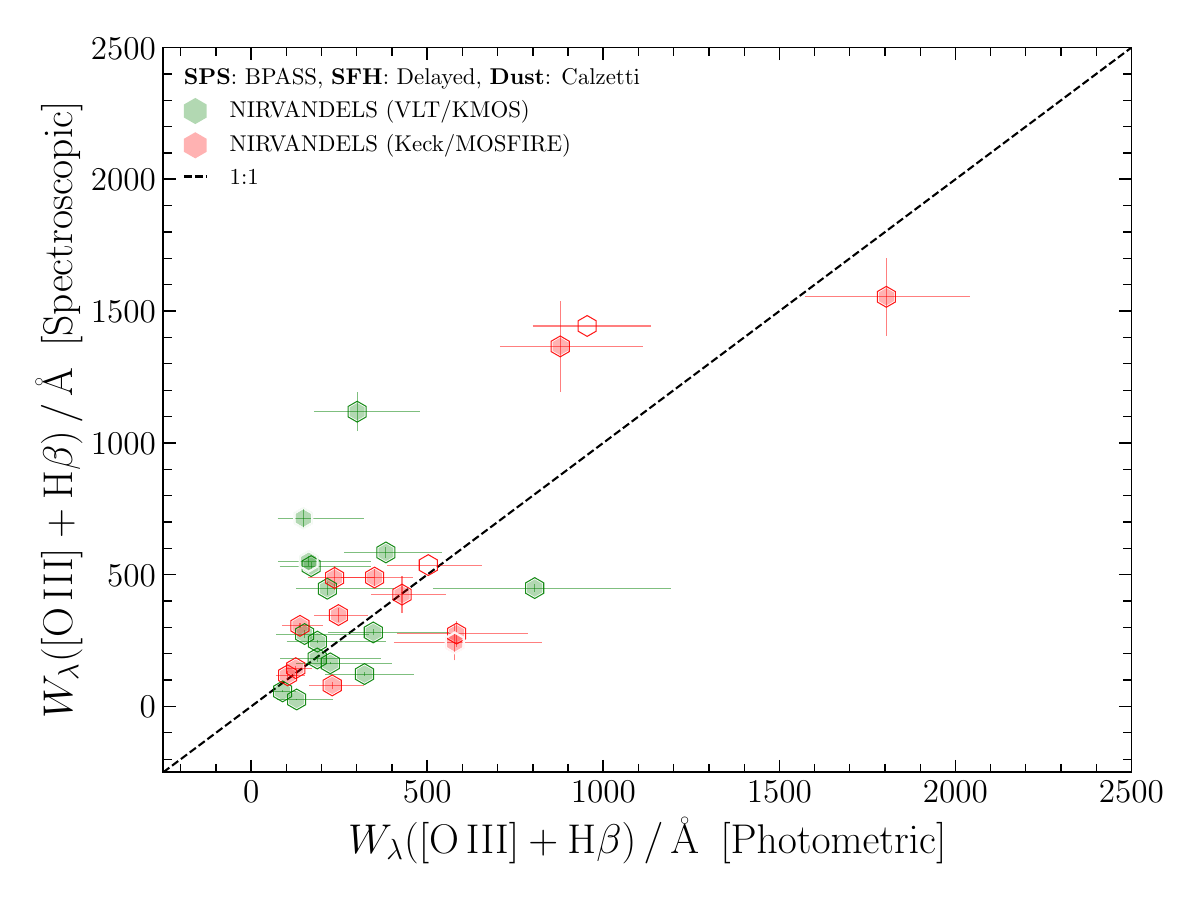}} \\ \subfloat[Delayed$-\tau$ SFH, SMC dust attenuation law, BPASS SPS models]{\includegraphics[width=1.0\columnwidth]{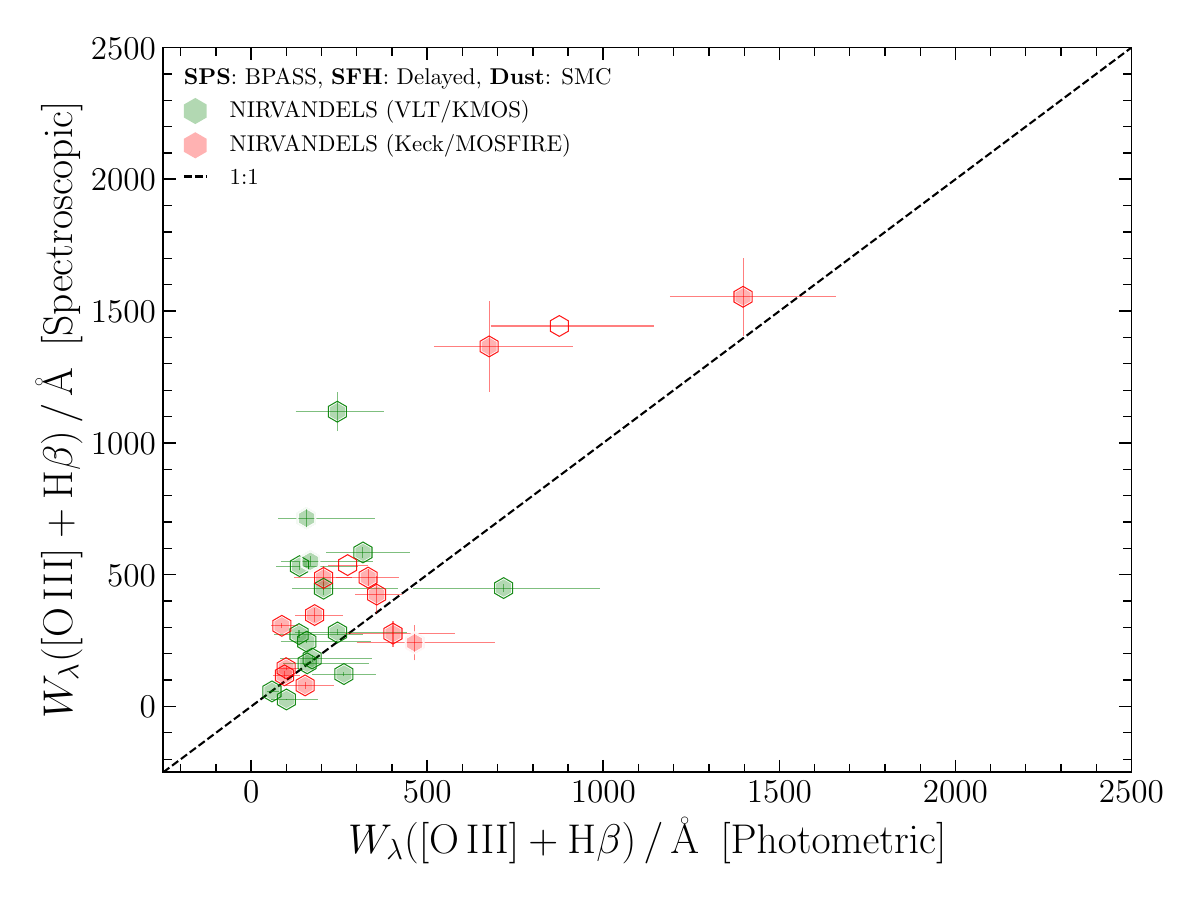}} & \subfloat[Continuity SFH, Calzetti dust attenuation law, BC03 SPS models]{\includegraphics[width=1.0\columnwidth]{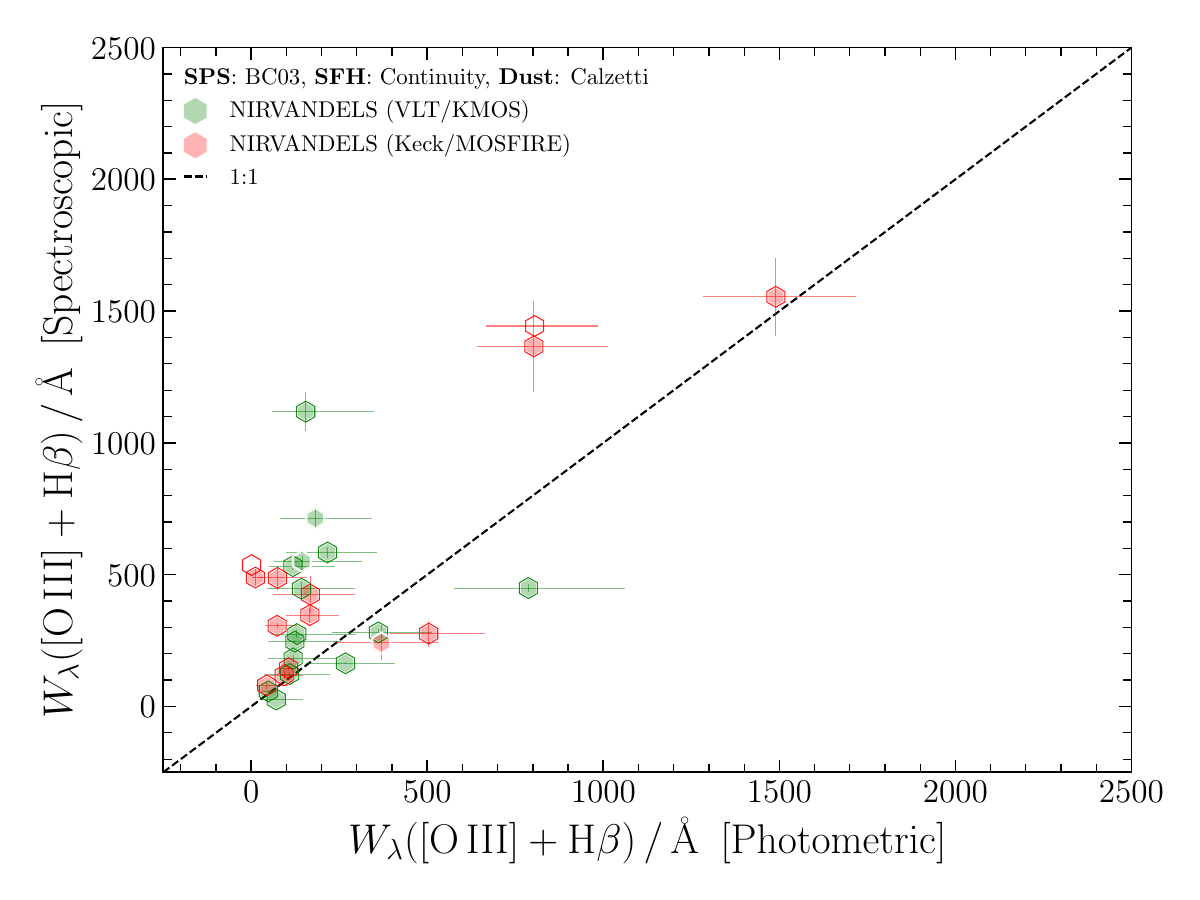}} 
    \end{tabular}
    \caption{Performance of different \textsc{bagpipes} configurations in recovering spectroscopic {\EWoiiihb} values from photometric SED fits for the NIRVANDELS spectroscopic sample.}
    \label{fig:appendix_A_nirv}
\end{figure*}


\bsp	
\label{lastpage}
\end{document}